\RequirePackage[switch, columnwise, running, mathlines, displaymath, mathlines]{lineno}
\documentclass[twocolumn,nofootinbib,superscriptaddress]{revtex4-2}

\usepackage[utf8]{inputenc}
\usepackage{amsmath}
\usepackage{gensymb}
\usepackage{amssymb}
\usepackage{graphicx}
\usepackage[dvipsnames]{xcolor}
\usepackage{hyperref}
\usepackage[normalem]{ulem}
\usepackage{aas_macros}
\usepackage{bm}
\usepackage{savesym}
\savesymbol{tablenum}
\usepackage{siunitx}
\restoresymbol{SIX}{tablenum}
\usepackage{enumitem}

\usepackage{tabularx}
\usepackage{colortbl}

\makeatletter
    \def\CT@@do@color{%
      \global\let\CT@do@color\relax
            \@tempdima\wd\z@
            \advance\@tempdima\@tempdimb
            \advance\@tempdima\@tempdimc
    \advance\@tempdimb\tabcolsep
    \advance\@tempdimc\tabcolsep
    \advance\@tempdima2\tabcolsep
            \kern-\@tempdimb
            \leaders\vrule
                    \hskip\@tempdima\@plus  1fill
            \kern-\@tempdimc
            \hskip-\wd\z@ \@plus -1fill }
\makeatother
\newcommand{\smu}{Department of Physics,
Southern Methodist University, 
Dallas, TX 75275, USA}

\newcommand{\Dunlap}{Dunlap Institute for Astronomy \& Astrophysics,
University of Toronto,
Toronto, ON M5S 3H4, Canada}

\newcommand{\UofT}{David A. Dunlap Department of Astronomy \& Astrophysics, 
University of Toronto,
Toronto, ON M5S 3H4, Canada}

\newcommand{\ASU}{School of Earth and Space Exploration,
Arizona State University, 
Tempe, AZ 85287, USA}

\def\vlo{\boldsymbol{\ell}_1}
\def\vlop{\boldsymbol{\ell}_1'}
\def\vlt{\boldsymbol{\ell}_2}
\def\vltp{\boldsymbol{\ell}_2'}
\def\vlA{\boldsymbol{\ell}_A}
\def\vlB{\boldsymbol{\ell}_B}
\def\vbL{\bm{L}}
\def\vlL{\boldsymbol{\ell}_L}
\def\lL{\ell_L}
\def\vlS{\boldsymbol{\ell}_S}
\def\lS{\ell_S}
\def\vL{\bm{\check{L}}}

\mathchardef\mhyphen="2D
\def\planck{\textit{Planck}}
\def\dd{{\rm d}}

\def\Lcheck{{\check{L}}}

\def\ALv{A_\Lcheck}
\def\PsiLv{\Psi_\Lcheck}
\def\ClTT{C_\ell^{TT}}
\def\lClTT{\tilde{C}_\ell^{TT}}
\def\ClTTobs{\hat{C}_\ell^{TT}}
\def\CLkk{C_L^{\kappa\kappa}}
\def\CLkkobs{\hat{C}_L^{\kappa\kappa}}
\def\NLkk{N_L^{(1),\kappa\kappa}}
\def\RDNo{N_L^{(0),\kappa\kappa,\mathrm{RD}}}
\def\Nokk{N_L^{(0),\kappa\kappa}}
\def\K{\mathrm{K}}
\def\k{\mathrm{k}}
\def\eqnname{Eq.}

\definecolor{purple}{rgb}{0.7, 0.2, 0.7}
\definecolor{red}{rgb}{0.7,0.1,0.1}
\definecolor{blue}{rgb}{0.2,0.4,0.8}
\definecolor{green}{rgb}{0.15,0.7,0.2}

\begin{document}

\title{Running the small-correlated-against-large estimator at scale: Applications of small-scale CMB lensing estimators on realistic simulations}

\author{Victor C. Chan}
\affiliation{\UofT}
\affiliation{\smu}

\author{Ren\'ee Hlo\v{z}ek} 
\affiliation{\Dunlap}
\affiliation{\UofT}

\author{Joel Meyers}
\affiliation{\smu}

\author{Alexander van Engelen}
\affiliation{\ASU}
\begin{abstract}
    The Small-Correlated-Against-Large Estimator (SCALE) for small-scale lensing of the cosmic microwave background (CMB) provides a novel method for measuring the amplitude of CMB lensing power without the need for reconstruction of the lensing field. In our previous study, we showed that the SCALE method can outperform existing reconstruction methods to detect the presence of lensing at small scales ($\ell \gg 3000$). Here we develop a procedure to include information from SCALE in cosmological parameter inference. We construct a precise neural network emulator to quickly map cosmological parameters to desired CMB observables such as temperature and lensing power spectra and SCALE cross spectra. We also outline a method to apply SCALE to full-sky maps of the CMB temperature field, and construct a likelihood for the application of SCALE in parameter estimation. SCALE supplements conventional observables such as the CMB power spectra and baryon acoustic oscillations in constraining parameters that are sensitive to the small-scale lensing amplitude such as the neutrino mass $m_\nu$. We show that including estimates of the small-scale lensing amplitude from SCALE in such an analysis provides enough constraining information to measure the minimum neutrino mass at $4\sigma$ significance in the scenario of minimal mass, and higher significance for higher mass.  Finally, we show that SCALE will play a powerful role in constraining models of clustering that generate scale-dependent modulation to the distribution of matter and the lensing power spectrum, as predicted by models of warm or fuzzy dark matter.
\end{abstract}
\maketitle
\section{Introduction}\label{sec:Intro}
Forthcoming high-resolution, low-noise observations of the cosmic microwave background (CMB) will allow analysis of its gravitational lensing features to greatly exceed the precision of current measurements.  Gravitational lensing is a particularly useful probe of cosmological density fluctuations, as it provides an unbiased tracer of the total mass density.  CMB lensing, in particular, utilizes a very well-characterized source plane at a known redshift, and thereby is especially powerful in this regard.

\begin{figure}[h!]
    \centering
    \includegraphics[width=\hsize]{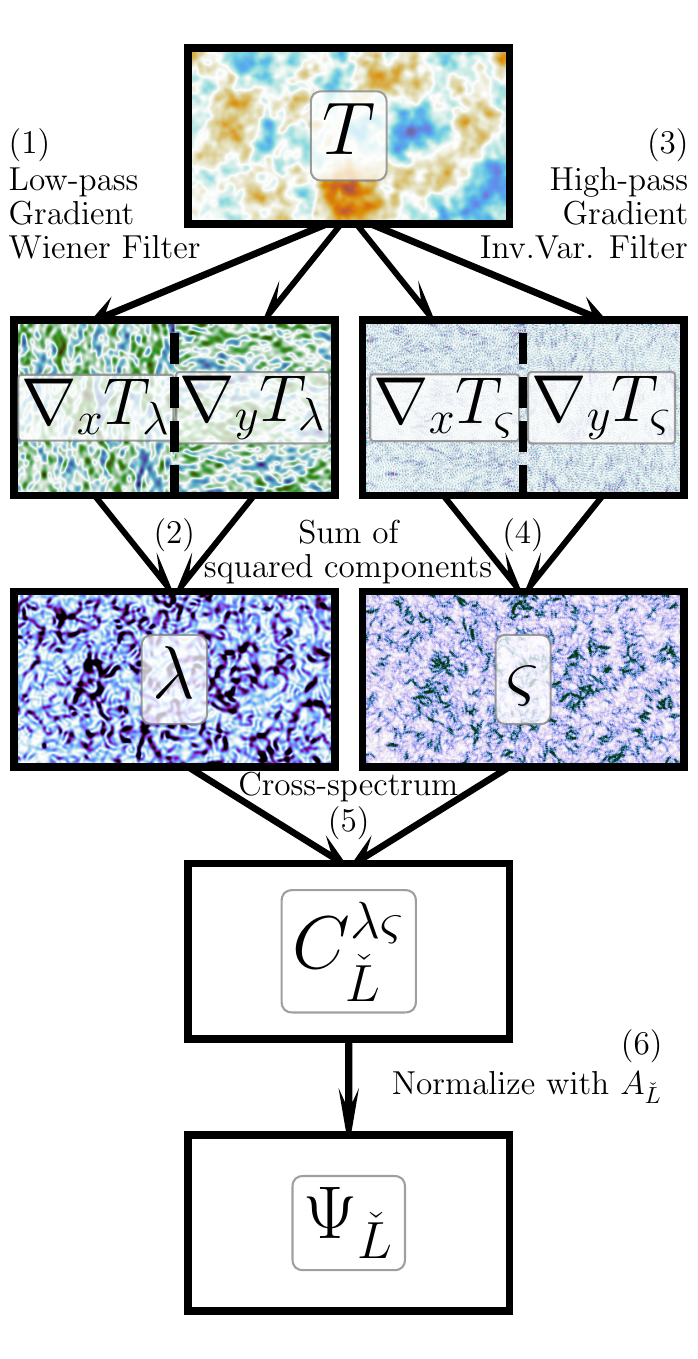}
    \caption{Summary of the steps involved in the Small-Correlated-Against-Large Estimator for small-scale CMB lensing. Small-scale features $\varsigma$ are correlated against large-scale features $\lambda$ which are dominated by the primary signal. 
    The presence of lensing at small-scales necessarily generates a positive cross-spectrum $C_\Lcheck^{\lambda\varsigma}$.}
    \label{fig:SCALEsteps}
\end{figure}

In Ref.~\citep[][hereafter C24]{Chan:2024}, we developed the Small-Correlated-Against-Large Estimator (SCALE) for measuring the small-scale ($\ell \gg 3000$) lensing power in the CMB. We showed that at the sensitivity of upcoming CMB experiments, SCALE can exceed the signal-to-noise with which small-scale lensing can be measured when compared with lensing reconstruction based on the standard quadratic estimator (QE)~\cite{Hu:2001tn,Hu:2001kj,Okamoto:2003zw}.
SCALE is a novel method to quantify small-scale CMB lensing in high-resolution temperature maps.
The dominant contribution to the CMB temperature power on small angular scales comes from lensing, and the amplitude of the lensing-induced power is directly proportional to the primary CMB gradient power on larger scales ($\ell\lesssim2000$)~\cite{Lewis:2006fu,Hadzhiyska:2019cle,Chan:2024}. An effective estimator can be constructed by taking advantage of this relationship. Furthermore, other features present at these scales (such as contributions to the observed intensity from astrophysical foregrounds) do not share the same direct correlations.
% \vinl{CMB lensing is the dominant signal at these angular scales, and it is strongly dependent on (directly tied to?) the gradient of the primary CMB at larger scales \joel{and the amplitude of lensing-induced small-scale fluctuations is directly proportional to the primary CMB gradient power on larger scales} \alex{"dominant CMB signal", not "dominant signal" (as that would be radio gals or dusty gals.)} ($\ell \lesssim 2000$) \cite{Lewis:2006fu,Hadzhiyska:2019cle,Chan:2024}. An effective estimator can be constructed by taking advantage of this principle \joel{relationship rather than principle?}, and the fact that other features present at these scales do not share the same direct correlations.} \victor{Second sentence may be too strong without proof that we're currently studying, suggestions welcome.} \joel{I think the second sentence is fine for an introduction.  This spells out the motivation, even if one could find counter-examples that make it not true in complete generality.}
The basic flow of the SCALE method is to pre-process a temperature map into a map $\lambda$ of the large-scale ($\lL < 3000$) gradient power which is dominated by primary CMB features, and a map $\varsigma$ of small-scale ($\lS \gg 3000$) gradient power which is dominated by lensing features. The normalized cross-spectrum of these two maps $\Psi_\Lcheck$ (defined later in \S~\ref{sec:SCALEApp/Emulator}) is directly related to the amplitude of the lensing power spectrum $\CLkk$ of the small-scale regimes associated with $\varsigma$ (see also Ref.~\cite{Zaldarriaga:2000} for a similar construction). 
These steps are summarized in \figurename~\ref{fig:SCALEsteps}. 
We refer the reader to C24 for a complete description of the principles of SCALE as well as comparisons to measurements with quadratic estimator reconstructions \cite{Hu:2001kj,Hu:2007bt}. Throughout the paper, we will typically use $\ell$ to index multipole moments of the CMB temperature, $L$ for those of the CMB lensing field, and $\Lcheck$ for the SCALE cross-spectra.

The value added by using SCALE compared to only utilizing conventional quadratic estimators lies in a few key areas. First, it is a simple method to quickly measure the amplitude of the small-scale CMB lensing power spectrum without the need for a full reconstruction of the lensing field. We established in C24 that SCALE outperforms quadratic estimators in terms of signal-to-noise of recovered lensing signal at small-scales in upcoming experiments. 
Quadratic estimators remain highly effective, tested, and well-understood tools for estimating CMB lensing on larger scales \cite{Planck:2015mym,Wu:2019hek,Planck:2019nip,ACT:2023kun,ACT:2023dou}, while SCALE is presented with an opportunistic, complementary role in the realm of small-scale lensing. If one wishes to use quadratic estimators at small-scales, the reconstruction bias $\NLkk$ grows to similar amplitude as the lensing power spectrum $C_L^{\kappa\kappa}$ by $L \sim 1000$, and dominates at higher $L$ \cite{Kesden:2003,Hanson:2011}. 
The small-scale lensing regime will become ever-more important as the experimental sensitivity improves, with surveys like the upcoming Simons Observatory (SO) \cite{SimonsObservatory:2018koc} and CMB-S4 \cite{CMB-S4:2016ple,Abazajian:2019eic}, as well as proposed future surveys like CMB-HD~\cite{Sehgal:2019ewc}. Improvements in foreground characterization and mitigation also suggest that we will be able to extract a significant amount of cosmology on these scales with appropriate statistical estimators. 

The maximum likelihood~\cite{Hirata:2002jy}, maximum a posteriori~\cite{Carron:2017mqf,Legrand:2023}, gradient inversion~\cite{Hadzhiyska:2019cle}, and Bayesian techniques~\cite{ Millea:2020cpw,Millea:2020iuw,Millea:2021had}  
%\alex{Put citations one by one, I think, not in a block at the end} 
are also currently being developed with the aim to tackle these challenges with small-scale lensing reconstruction and go beyond what the QE is capable of.  
%\alex{Need a sentence and cite for the SPT MUSE paper (Ge+) which is now one of the leading CMB lensing measures. } 
For example, Bayesian lensing techniques aim to jointly provide a probabilistic estimate of cosmological parameters alongside a lensing potential map and a delensed primary CMB map~\cite{Millea:2017fyd,Millea:2020cpw,Millea:2021had}; these have recently been applied to data collected by SPTpol~\cite{Millea:2020iuw} and SPT-3G~\cite{SPT-3G:2024atg}. 
The maximum likelihood approach is capable of optimal estimation of the lensing field~\cite{Hirata:2002jy}, though there is a non-trivial computational cost involved with the iterative reconstruction it employs.
The maximum a posteriori method achieves nearly optimal reconstruction with an alternate iterative scheme~\cite{Carron:2017mqf,Legrand:2023}.
%\alex{What about MAP?} 
The gradient inversion estimator aims to improve small-scale lensing reconstruction by introducing normalization that is dependent on the local temperature gradient~\cite{Hadzhiyska:2019cle}.  Each of these methods achieves sensitivity to lensing exceeding the quadratic estimator, for low-noise data.   SCALE shares the same guiding principle of these beyond-QE estimators, by correlating small-scale lensing features with the background temperature gradient. In particular SCALE shares the advantage of the gradient inversion estimator by avoiding estimator variance due to the background gradient, but it achieves this using explicit correlation with the gradient, rather than direct inversion.  

Reconstructions of the small-scale lensing potential field are desirable for cross-correlation with other probes of the distribution of matter density, but SCALE excels in its simplicity and speed in extracting small-scale lensing information from a CMB map. Although SCALE operates within a similar regime as these other techniques, it remains unique by entirely bypassing the need for map-level reconstruction in order to estimate small-scale lensing information. This quality makes it difficult for the performance of SCALE to be directly compared with any small-scale lensing reconstruction methods, but a fair comparison can be made at the likelihood level. It is also possible that SCALE takes advantage of contractions in the CMB lensing four-point function, or trispectrum, to which map-based estimators are insensitive.

Small-scale lensing of the CMB is a particularly interesting laboratory for probing the dark matter and clustering properties of the universe (see Ref.~\cite{Lewis:2006fu} for a review of CMB lensing). The effect of massive neutrinos on the small-scale lensing power spectrum is a nearly scale-independent suppression, and an accurate measurement of the small-scale lensing amplitude will provide a valuable probe of the total neutrino mass~\cite{Kaplinghat:2003bh,Lesgourgues:2006nd}.  Cosmological measurement of neutrino mass is complementary to lab-based probes of neutrino mass~\cite{Gerbino:2022nvz} and allows for insights into physics beyond the Standard Model~\cite{Dvorkin:2019jgs,Green:2021xzn}.  CMB lensing probes of neutrino mass have recently taken on additional importance given hints of tighter than expected upper bounds on neutrino mass with existing cosmological data~\cite{Craig:2024tky,Wang:2024hen,Green:2024xbb,Naredo-Tuero:2024sgf,Jiang:2024viw}. 
Some non-standard models of warm or fuzzy dark matter also predict a suppression of clustering on small-scales, which in turn leads to a phenomenological, scale-dependent suppression of lensing power \cite{Primack:2001,Palanque-Delabrouille:2019iyz,Lovell:2014,Hui:2017,Hlozek:2018,Lague:2022,Dentler:2022}. These scale-dependent effects could be constrained with multiple measurements of the lensing amplitude at different scales~\cite{Nguyen:2017zqu,MacInnis:2024znd}.

In this work, we build on the established SCALE method, and extend its application to full-sky maps. We also study its constraining power when applied to practical cosmological parameter estimation, particularly in the context of small-scale lensing. We do not include a characterization of CMB foregrounds in this work, as small-scale power added by foregrounds is not expected to correlate with the large-scale features of the primary CMB temperature field; this is still an interesting hypothesis to test, and we will leave further study of foregrounds for future work. Sampling cosmological parameters quickly requires fast predictions of theoretical spectra to compare to within the likelihood. We construct a neural network emulator and present its performance in \S\ref{sec:SCALEApp/Emulator}. The emulator provides significant speed benefits when mapping a set of cosmological parameters to expected band-powers of angular power spectra with only a small penalty to precision ($\lesssim 0.5\%$). We present our suite of full-sky simulations for the lensed CMB in \S\ref{sec:SCALEApp/Sims}. We construct a likelihood for CMB temperature power spectra in a $\Lambda$CDM parameter estimation, and then extend the model to include conventional lensing reconstruction information as well as SCALE cross spectra in \S\ref{sec:SCALEApp/Likelihood}, and present the results in \S\ref{sec:SCALEApp/Results}. We additionally include an analysis with a model including a general, scale-dependent suppression of lensing at extremely small scales ($L \sim 10\,000$), and we show that SCALE can be configured to accurately detect exotic models of dark matter. Finally, we discuss the results and conclude in \S\ref{sec:SCALEApp/Concl}.
\section{Emulation of CMB and SCALE spectra}\label{sec:SCALEApp/Emulator}
We begin with the development of a set of emulators to quickly predict theoretical observables including the CMB lensed temperature power spectra $\ClTT$, lensing power spectra $\CLkk$, lensing reconstruction bias $\NLkk$, as well as analytic SCALE products $\ALv$ and $\PsiLv$ (whose values are defined below in Eqs.~\eqref{eq:SCALEApp/ALDef} and \eqref{eq:SCALEApp/PsiExpected}). The posterior sampling process typically requires upwards of $\mathcal{O}(10^3$ to $10^4)$ steps per chain in order to reasonably explore a hyperspace of several cosmological parameters, and that necessitates a quick mapping from the set of parameters at each step to the relevant observables. 
While CMB angular power spectra can be computed quickly with existing software, repeatedly computing the expected SCALE cross-spectra with Eqs.~\eqref{eq:SCALEApp/ALDef} and \eqref{eq:SCALEApp/PsiExpected} from different sets of parameters requires a significant speedup over conventional numerical integration methods.\footnote{All computation speeds reported in this work are timed with an AMD Ryzen 9 5900X CPU with 12 physical cores and 24 logical cores.}
We show that neural network (NN) emulators can predict SCALE observables at the speed required for quick posterior sampling without a significant penalty in terms of accuracy. The emulators can also be trained to predict lensed CMB temperature power spectra $\lClTT$ and lensing power spectra $\CLkk$ faster than Boltzmann codes, while maintaining high accuracy.
\subsection{Timing of calculations without emulators}\label{sec:SCALEApp/Emulator/Pretiming}
Given a set of cosmological parameters, there are now Boltzmann codes that quickly and accurately compute primary CMB power spectra $\ClTT$, lensing power spectra $C_L^{\phi\phi}$, and lensed CMB power spectra $\lClTT$ (\texttt{CAMB}\footnote{\url{https://camb.info/}}, \cite{CAMB:2011}; \texttt{CLASS}\footnote{\url{https://github.com/lesgourg/class_public}}, \cite{CLASS:2011b}). We opt to use \texttt{CAMB} in this work to keep consistency with C24, but the applications should be comparable to outputs from \texttt{CLASS}. Consider computing example power spectra in \texttt{CAMB} with \texttt{lens\_potential\_accuracy=8} as the only non-default parameter to ensure lensing accuracy at high-$\ell$ \cite{McCarthy:2022}. \texttt{CAMB} is able to compute and return the power spectra in $\mathcal{O}(1\,{\rm s})$, with some mild dependence on the requested \texttt{lmax}. This is fast enough to be used in parameter estimation, but it is possible to speed up the process further by using an emulator. This is especially true if one is interested in running modified versions such as \texttt{axionCAMB}\footnote{\url{https://github.com/dgrin1/axionCAMB}} which take extra computational steps to include non-standard physics which could affect the small-scale lensing power spectrum and can extend the computational time by factors of anywhere from two to ten \cite{axionCAMB:2022}.

A stronger motivation for the construction of emulators for theoretical spectra comes from the application of SCALE in a cosmological likelihood. The analytic forms for SCALE products were presented in C24, and are repeated here:
\begin{align}\label{eq:SCALEApp/ALDef}
    &A_\Lcheck = \Bigg[ 2 \int \frac{\dd^2\vlS}{(2\pi)^2} W_\varsigma(\vlS)W_\varsigma(\vL-\vlS)
    \nonumber \\
    & \qquad\qquad \times \left(\vlS \cdot (\vlS-\vL) \right) \frac{1}{C_{\lS}^{TT,\mathrm{obs}}} \frac{1}{C_{|\vL-\vlS|}^{TT,\mathrm{obs}}}   
    \nonumber \\
    & \qquad\qquad \times 
    \int \frac{\dd^2\vlL}{(2\pi)^2} W_\lambda(\vlL)W_\lambda(\vL-\vlL)
    \left(\vlL \cdot (\vlL-\vlS) \right) \nonumber \\
    & \qquad\qquad\quad \times \left( (\vL-\vlL) \cdot (\vlS-\vlL) \right) \left(\vlL \cdot (\vlL-\vL) \right) \nonumber \\
    & \qquad\qquad\quad \times \frac{C_{\lL}^{TT} C_{\lL}^{TT,\mathrm{fid}}}{C_{\lL}^{TT,\mathrm{obs}}} \frac{C_{|\vL-\vlL|}^{TT} C_{|\vL-\vlL|}^{TT,\mathrm{fid}}} {C_{|\vL-\vlL|}^{TT,\mathrm{obs}}} \Bigg]^{-1} \, , 
\end{align}
\begin{align}\label{eq:SCALEApp/PsiExpected}
    &\left\langle \Psi_\Lcheck \right\rangle = 2 A_\Lcheck
    \int \frac{\dd^2\vlS}{(2\pi)^2} W_\varsigma(\vlS)W_\varsigma(\vL-\vlS) \nonumber \\
    & \qquad\qquad \times \left(\vlS \cdot (\vlS-\vL) \right) \frac{1}{C_{\lS}^{TT,\mathrm{obs}}} \frac{1}{C_{|\vL-\vlS|}^{TT,\mathrm{obs}}} \nonumber \\
    & \qquad\qquad \times 
    \int \frac{\dd^2\vlL}{(2\pi)^2} W_\lambda(\vlL)W_\lambda(\vL-\vlL)
    \left(\vlL \cdot (\vlL-\vlS) \right)
     \nonumber \\
    & \qquad\qquad\quad \times \left( (\vL-\vlL) \cdot (\vlS-\vlL) \right)
    \left(\vlL \cdot (\vlL-\vL) \right) \nonumber \\
    & \qquad\qquad\quad \times \frac{C_{\lL}^{TT} C_{\lL}^{TT,\mathrm{fid}}}{C_{\lL}^{TT,\mathrm{obs}}} \frac{C_{|\vL-\vlL|}^{TT} C_{|\vL-\vlL|}^{TT,\mathrm{fid}}} {C_{|\vL-\vlL|}^{TT,\mathrm{obs}}} C_{|\vlS-\vlL|}^{\phi\phi} \, ,
\end{align}
where $A_\Lcheck$ is the normalization for a cross spectrum $C_\Lcheck^{\lambda\varsigma}$ between \textit{large-scale} gradient power $\lambda$ and \textit{small-scale} gradient power $\varsigma$ such that $\Psi_\Lcheck = A_\Lcheck C_\Lcheck^{\lambda\varsigma}$. 
These equations deviate slightly from their presentation in C24; we generalize them here to include the expected SCALE response with respect to changes in cosmology (reflected in the expected primary temperature power $C_\ell^{TT}$) while holding our choice in filters fixed (reflected by the total observed power $C_\ell^{TT,\mathrm{obs}} = \tilde{C}_\ell^{TT,\mathrm{fid}} + N_\ell^{TT}$ and the expected primary temperature power at a fiducial cosmology $C_\ell^{TT,\mathrm{fid}}$). 
These integrals are constructed in a 4-dimensional Fourier space, and an implementation in which the integrals are numerically computed with the mid-point rule is provided in our publicly available package \texttt{cmbpix}\footnote{\url{https://github.com/victorcchan/cmbpix}}. Consider an example computation of analytic SCALE $\PsiLv$ for multipoles $2 < \Lcheck < 2002$ such that the width of the small-scale window for $\varsigma$ is $\ell_{S,\mathrm{max}} - \ell_{S,\mathrm{min}} = 2000$, the large-scale window for $\lambda$ is $0 < \lL < 3000$, and each $\PsiLv$ is evaluated with a Riemann sum on 2-dimensional grids of $\Delta\lS=75$ and $\Delta\lL = 100$. This takes $\mathcal{O}(10^4\,{\rm s})$ or $\mathcal{O}(10\,{\rm min})$ to compute. The numerical accuracy and computational speed of \eqnname~\eqref{eq:SCALEApp/ALDef}-\eqref{eq:SCALEApp/PsiExpected} is dependent on the resolution of the grid(s) on which it is evaluated. Regardless, numerically integrating the analytic SCALE products with the mid-point rule is slow enough that one would desire considerable speedups. One approach to speeding up the evaluation of \eqnname~\eqref{eq:SCALEApp/ALDef}-\eqref{eq:SCALEApp/PsiExpected} is to consider Monte Carlo (MC) integration. We find that a Monte Carlo integration offers a good balance of speed and accuracy and is implemented in \texttt{cmbpix}. One drawback of Monte Carlo integration is its non-deterministic nature, manifesting as an inherent imprecision which is dependent on the number of samples with which the integral is evaluated. We find that MC integration of \eqnname\eqref{eq:SCALEApp/ALDef}-\eqref{eq:SCALEApp/PsiExpected} produces approximately $\sim 1\%$ scatter (68\% region) when evaluated with $N_\mathrm{samples}\sim \mathcal{O}(10^{5})$ samples. This is the accuracy for the evaluation at one $\Lcheck$ mode, and when we bin SCALE powers into bands of width $\Delta\Lcheck=71$, we expect the scatter to be reduced by a factor of $\sqrt{\Delta\Lcheck} \approx 8.4$. Evaluating \eqnname\eqref{eq:SCALEApp/ALDef}-\eqref{eq:SCALEApp/PsiExpected} with $N_\mathrm{samples} = 2\times10^5$ within the MC integration takes $\mathcal{O}(10\,{\rm s})$. This is a significant speed-up, but a further increase in speed is desired for parameter estimation.

Emulation is further motivated if one wishes to include conventional lensing reconstruction information ($L \lesssim 1250$) from a quadratic estimator. Typically, an `observed' lensing power spectrum $\CLkkobs$ estimated from a QE reconstruction requires the subtraction of biases $\Nokk$ and $\NLkk$ such that \cite{Hu:2001kj,Kesden:2003,Regan:2010,Hanson:2011,Planck:2015mym,Wu:2019hek,ACT:2023kun}
\begin{equation}\label{eq:CLkkQE}
    \CLkk \sim \CLkkobs - \Nokk - \NLkk.
\end{equation}
The zeroth-order reconstruction bias $\Nokk$ includes contributions from the disconnected four-point function which is non-zero even in the absence of lensing. It is, in practice, dependent on the particular observed realization of the CMB \cite{Schmittfull:2013}, and the realization-dependent $\RDNo$ can be estimated with a combination of the observed CMB temperature power spectrum $\ClTTobs$ of the particular realization along with the theoretical power spectrum $\ClTT$ used in the QE filters \cite{Regan:2010,Planck:2015mym,Wu:2019hek,Planck:2019nip,ACT:2023kun,ACT:2023dou}. The first-order reconstruction bias $\NLkk$ includes non-Gaussian contributions from the connected four-point function that are not included in the quadratic estimator. For a configuration with isotropic noise, it can be analytically computed with an integral constructed in 4-dimensional Fourier space similar to \eqnname~\eqref{eq:SCALEApp/ALDef}-\eqref{eq:SCALEApp/PsiExpected}. Two equivalent representations are derived in Refs.~\cite{Kesden:2003,Hanson:2011}. The version from Ref.~\cite{Kesden:2003} for the auto-correlation of the reconstructed lensing potential field $\hat{\phi}$ with the standard temperature-temperature $(TT)$ quadratic estimator to estimate the lensing potential power $\hat{C}_L^{\phi\phi}$ is repeated here:
\begin{align}
    N_{TT,TT}^{(1),\phi\phi}(L) &= \frac{A_{TT}^2(L)}{L^2} \int \frac{\mathrm{d}^2\vlo}{(2\pi)^2} \int \frac{\mathrm{d}^2\vlop}{(2\pi)^2} \nonumber\\
    \times & F_{TT}(\vlo,\vlt) F_{TT}(\vlop,\vltp) \nonumber\\
    \times & \Big\{ C_{|\vlo-\vlop|}^{\phi\phi}f_{TT}(-\vlo,\vlop)f_{TT}(-\vlt,\vltp) \nonumber\\
    & + C_{|\vlo-\vltp|}^{\phi\phi}f_{TT}(-\vlo,\vltp)f_{TT}(-\vlt,\vlop) \Big\}.\label{eq:N1Kesden}
\end{align}
The normalization $A_{TT}$ as well as the lensing weight functions $f_{TT}$ and $F_{TT}$ are the same as those presented in Ref.~\cite{Hu:2001kj,Kesden:2003} such that $\vbL = \vlo + \vlt = \vlop + \vltp$ (we are considering contributions to the connected four-point function). 
We repeat the definitions of $A_{TT}$, $f_{TT}$, and $F_{TT}$ here:
\begin{equation}\label{eq:QENorm}
    A_{TT}(L) = L^2 \Big[ \int \frac{\mathrm{d}^2\vlo}{(2\pi)^2} f_{TT}(\vlo,\vlt)F_{TT}(\vlo,\vlt) \Big],
\end{equation}
\begin{equation}\label{eq:QEf}
    f_{TT}(\vlA,\vlB) = C_{\ell_A}^{T\nabla T} (\mathbf{L} \cdot \vlA) + C_{\ell_B}^{T\nabla T} (\mathbf{L} \cdot \vlB),
\end{equation}
\begin{equation}\label{eq:QEF}
    F_{TT}(\vlA,\vlB) = \frac{f_{TT}(\vlA,\vlB)}{2 C_{\ell_A}^{TT,\mathrm{obs}} C_{\ell_B}^{TT,\mathrm{obs}}},
\end{equation}
where $\vlA$ and $\vlB$ can be substituted with any of $\{ \vlo, \vlt, \vlop, \vltp \}$. 
$f_{TT}$ is defined with respect to the expected lensing response $C_\ell^{T\nabla T}$ \cite{Lewis:2011} which is readily computed in \texttt{CAMB}, and it is entirely determined by the effects of lensing on the temperature field. $F_{TT}$ represents the filter used in the quadratic estimator, and it is defined with respect to the expected lensing response $C_\ell^{T\nabla T}$ weighted by the total observed power $C_\ell^{TT,\mathrm{obs}}$. 
This distinction is relevant for computing the expected bias spectrum for different sets of cosmological parameters while holding the filtering fixed at a fiducial cosmology. Repeatedly computing the expected $\NLkk$ during parameter estimation using \eqnname~\eqref{eq:N1Kesden} is prohibitively slow for the same reasons as for the SCALE observables. 
In practice, $\NLkk$ is typically estimated through Monte Carlo simulations of observational noise at a fiducial cosmology, and its dependence on cosmological parameters is usually included as an approximation in the likelihood with its dependence (both directly and indirectly) on $\ClTT$ and $\CLkk$ \cite{Planck:2015mym,Wu:2019hek,ACT:2023kun}. 
These approximations rely on first-order derivatives of Eq.~\eqref{eq:N1Kesden} with respect to $\ClTT$ and $\CLkk$ which require calculations similar to Eq.~\eqref{eq:N1Kesden}. 
An emulator can quickly predict $\NLkk$ as computed with Eq.~\eqref{eq:N1Kesden} for a wide range of cosmological parameters, which removes the need for its approximation. This approach works well for our simple models/analyses, though the Monte Carlo computation of $\NLkk$ and the associated approximations are well-suited for observed data which is contaminated with foregrounds, masking, etc. 
\subsection{Cosmological model}\label{sec:SCALEApp/Emulator/Model}
\begin{figure}
    \centering
    \includegraphics[width=0.95\hsize]{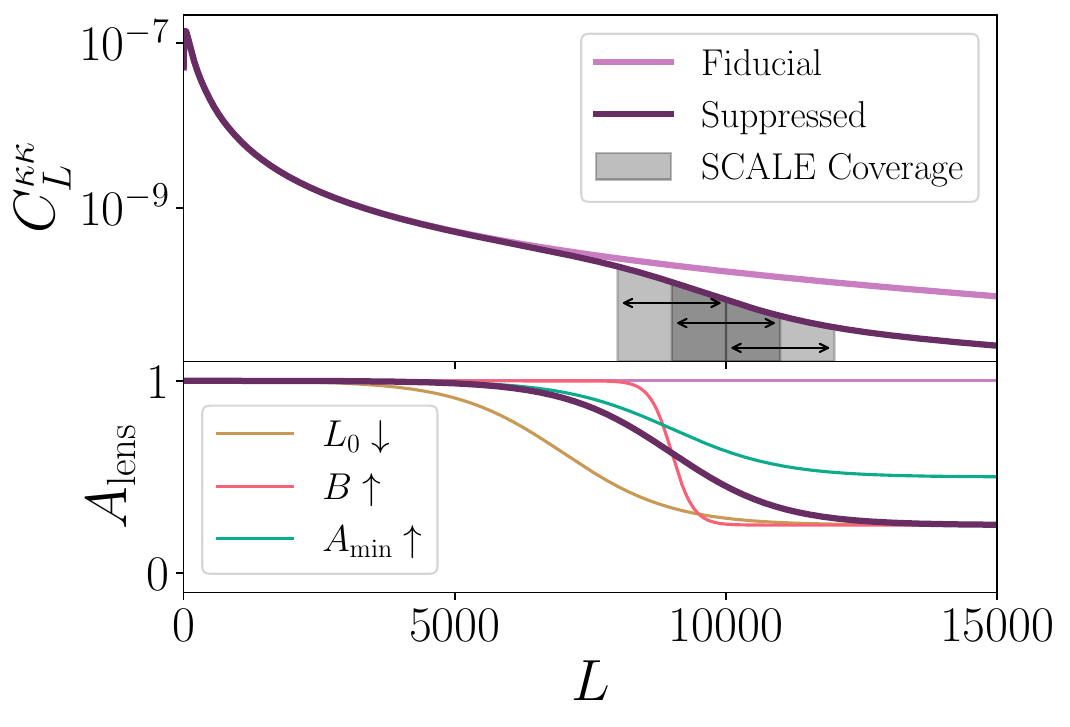}
    \caption[Visualization of the lensing suppression model]{\textit{Top}: The lensing convergence power spectrum $C_L^{\kappa\kappa}$ for the fiducial cosmology as computed from \texttt{CAMB} is compared to a model with a suppression of lensing power applied at small scales. Also shown are the three small-scale filter ranges for the applications of SCALE chosen to recover information about the lensing suppression. \textit{Bottom}: The lensing amplitude $A_{\rm lens}$ applied to $C_L^{\kappa\kappa}$ in our suppression model is shown in dark purple (see parameters in Table~\ref{tab:Fiducial}). A version with lower $L_0 = 7\,000$ is shown in yellow. A version with higher $B = 0.005$ is shown in red. A version with higher $A_{\rm min} = 0.5$ is shown in green. Also shown is the fiducial model with no suppression obtained by setting $A_{\rm min} = 1$.}
    \label{fig:LensSuppression}
\end{figure}

Consider a simple cosmological model for which conventional CMB observables can be computed with \texttt{CAMB} or \texttt{CLASS}, including the six $\Lambda$CDM parameters $\{ \Omega_c,\allowbreak\, \Omega_b,\allowbreak\, A_s,\allowbreak\, n_s,\allowbreak\, h,\allowbreak\, \tau \}$ and single massive neutrino species with mass $m_\nu$. The chosen fiducial values are listed in Table~\ref{tab:Fiducial}. 
This assumed base model is a good test of the power of SCALE to constrain the small-scale lensing power, which is particularly sensitive to $m_\nu$~\cite{Kaplinghat:2003bh,Lesgourgues:2006nd,Green:2021xzn}. 
\begin{table}
    \centering
    \caption[Set of fiducial cosmological parameters]{The set of cosmological parameters chosen for the fiducial model.}
    \label{tab:Fiducial}
    \begin{tabular}{|l|l|c|}
        \hline
        \rowcolor[HTML]{E7F9FF} Parameter & Symbol & Value \\
        \hline
        Reduced Hubble constant & $h$ & $0.675$ \\
        \rowcolor[HTML]{EFEFEF} Baryon density & $\Omega_\mathrm{b}h^2$ & $0.022$ \\
        (Cold) Dark matter density & $\Omega_\mathrm{c}h^2$ & $0.122$ \\
        \rowcolor[HTML]{EFEFEF} Optical depth to reionization & $\tau$ & $0.06$ \\
        Scalar fluctuation amplitude & $A_\mathrm{s}$ & $2.1 \times 10^{-9}$ \\
        \rowcolor[HTML]{EFEFEF} Scalar spectral index & $n_\mathrm{s}$ & $0.965$ \\
        Neutrino mass & $m_\nu$ & $0.06\,{\rm eV}$ \\
        \rowcolor[HTML]{EFEFEF} Lens suppression scale & $L_0$ & ${}^a\,9\,000$ \\
        Lens amplitude decay rate & $B$ & ${}^a\,0.001$ \\
        \rowcolor[HTML]{EFEFEF} Lens suppression depth & $A_{\rm min}$ & ${}^a\,1,\,{}^b\,(0.25)$ \\
        Experiment noise & $w$ & ${}^c 1\,\mu\K$-arcmin \\
        \rowcolor[HTML]{EFEFEF} Experiment beam & $\sigma_b$ & ${}^c 1\,{\rm arcmin}$ \\
        \hline
    \end{tabular}\\
    {\footnotesize
    ${}^a$Lensing suppression parameters are held fixed at these values for base analysis, with $A_{\rm min}=1$ resulting in effectively no suppression.\\
    ${}^b$ A separate simulation with $A_{\rm min}=0.25$ was generated for analysis including lensing suppression.\\
    ${}^c$Experimental configuration similar to a CMB-S4-like survey chosen to match the analysis of Configuration D in C24. 
    }
\end{table}
In addition to this base model where the lensing power spectrum is largely scaled by $m_\nu$, we consider a phenomenological model for scale-dependent lensing power suppression that is a feature of many dark matter models beyond the standard picture. These models are easily modeled with \texttt{CAMB}'s included \texttt{get\_partially\_lensed\_cls} method, which applies a lensing amplitude function $A_{\rm lens}(L)$ to the non-suppressed lensing power $C_L^{\kappa\kappa}$ such that the suppressed lensing power spectrum $C_L^{\kappa\kappa,\mathrm{sup}}$ is
\begin{equation}\label{eq:LensSuppression}
    C_L^{\kappa\kappa,\mathrm{sup}} = A_{\rm lens}(L) C_L^{\kappa\kappa}.
\end{equation}
We choose to model the suppression of small-scale lensing using a function similar to a logistic function, parameterized by a suppression scale $L_0$, lensing amplitude decay rate $B$, and suppression amplitude $A_{\rm min}$:
\begin{equation}\label{eq:Alens}
    A_{\rm lens}(L) = \frac{1 - A_{\rm min}}{1 + \exp\left(B(L-L_0)\right)} + A_{\rm min}.
\end{equation}
This function asymptotes to unity for $L \ll L_0$, and to $A_{\rm min}$ for $L \gg L_0$, and the decay rate parameter $B$ sets the steepness of the transition. The shape of the suppression model is reminiscent of those predicted by fuzzy dark matter models as shown in e.g. Ref.~\cite{Hlozek:2018}. Our choice in the fiducial values for the suppression model are listed in Table~\ref{tab:Fiducial}, and the corresponding $A_{\rm lens}$ is illustrated in \figurename~\ref{fig:LensSuppression}. This specific model exhibits a modulation in the CMB lensing power at small-scales where SCALE is particularly effective. 
Our emulator is capable of handling input lensing suppression parameters in each prediction, but in our base analysis we fix $A_{\rm min} = 1$ to achieve effectively no suppression for all $L$. In other words, we predict power spectra with standard lensing by passing in $A_{\rm min} = 1$ into our emulator. We later explicitly turn on lensing suppression for a separate analysis by including $A_{\rm min}$ as a free parameter, and generating a new simulation with an underlying value of $A_{\rm min} = 0.25$, which we describe later. 
A visualization of the chosen suppression model is shown in Figure~\ref{fig:LensSuppression}, along with a few examples where the parameters $L_0$, $
B$, and $A_{\rm min}$ are varied individually.
\subsection{Construction of emulators}\label{sec:SCALEApp/Emulator/Emulator}
We construct a single emulator to predict theoretical CMB (partially) lensed $TT$ power spectra $\tilde{C}_\ell^{TT}$, lensing convergence power spectra $C_L^{\kappa\kappa}$, lensing reconstruction bias $\NLkk$, and SCALE spectra $\Psi_\Lcheck$ for a wide range of cosmological parameters. The emulators are created in the \texttt{COMSOPOWER}\footnote{\url{https://github.com/alessiospuriomancini/cosmopower}} framework \cite{COSMOPOWER:2022}, which is a Python package for the construction of emulators for cosmological observables. It is based on \texttt{TensorFlow}\footnote{\url{https://www.tensorflow.org/}} \cite{abadi2016tensorflow}, and it provides a structure for training and using neural network (NN) emulators.
\subsubsection{Training data}\label{sec:SCALEApp/Emulator/Emulator/Training}
The training data consist of $N_\mathrm{train}=8192$ sets of cosmological parameters sampled from a Latin hypercube for uniform priors in the ranges set by Table~\ref{tab:TrainingPriors}. 
This is a relatively small number of training points compared to a characteristic $N_\mathrm{train}=\mathcal{O}(10^5)$ \cite{COSMOPOWER:2022}, but we show later that it is sufficient to train the emulators to a satisfactory degree of accuracy. 
The range of $\Lambda$CDM parameters is chosen to be centered on the reported Planck 2018 values with width $\pm 4.5\sigma$ \cite{Planck:2018vyg}. We choose a training range of $[0,2]$ for the lensing suppression parameter $A_{\rm min}$, to ensure that the emulator is well-trained to recover the cosmology in the `vanilla' case where  $A_{\rm min} = 1$ and thus where the suppression is effectively turned off as well as the case where suppression is explicitly included.
\begin{table}
    \centering
    \caption[Prior ranges for emulator training data]{The set of cosmological parameters and their prior ranges used to generate training data for the emulators. The $\Lambda$CDM parameters are chosen to be centered on the Planck 2018 best-fit cosmology with $\pm 4.5\sigma$ on either side \cite{Planck:2018vyg}. Note that the emulators are trained on parameters $\Omega_\mathrm{b}$, $\Omega_\mathrm{c}$, and $\ln(10^{10}A_s)$, but are sampled as shown as direct inputs for \texttt{CAMB}.}
    \label{tab:TrainingPriors}
    \begin{tabular}{|l|l|c|}
        \hline
        \rowcolor[HTML]{E7F9FF} Parameter & Symbol & Prior range \\
        \hline
        Reduced Hubble constant & $h$ & $[0.6493, 0.6979]$ \\
        \rowcolor[HTML]{EFEFEF} Baryon density & $\Omega_\mathrm{b}h^2$ & $[0.021695, 0.023045]$ \\
        (Cold) Dark matter density & $\Omega_\mathrm{c}h^2$ & $[0.1146, 0.1254]$ \\
        \rowcolor[HTML]{EFEFEF} Optical depth to reionization & $\tau$ & $[0.02155, 0.08725]$ \\
        Scalar fluctuation amplitude & $A_\mathrm{s}$ & $[1.965, 2.235] \times 10^{-9}$ \\
        \rowcolor[HTML]{EFEFEF} Scalar spectral index & $n_\mathrm{s}$ & $[0.946, 0.9838]$ \\
        Neutrino mass & $m_\nu$ & $[0, 0.18]\,{\rm eV}$ \\
        \rowcolor[HTML]{EFEFEF} Lens suppression scale & $L_0$ & $[4\,000,16\,000]$ \\
        Lens amplitude decay rate & $B$ & $[10^{-4}, 0.05]$ \\
        \rowcolor[HTML]{EFEFEF} Lens suppression depth& $A_{\rm min}$ & [0,2]\\
        Small-scale filter width${}^a$ & $\Delta\lS$ & $[800, 3200]$ \\
        \rowcolor[HTML]{EFEFEF} Small-scale filter center${}^a$ &  $\overline{\lS}$ & $[6200, 9800]$ \\
        \hline
    \end{tabular}\\
    {\footnotesize
    ${}^a$Only affects SCALE band-powers.
    }
\end{table}
For each set of parameters in our training space, we compute the following:
\begin{enumerate}
    \item The unlensed $TT$ power spectrum $C_\ell^{TT}$ from \texttt{CAMB} for $2 \leq \ell \leq 20\,000$.
    \item The lensing response $C_\ell^{T\nabla T}$ from \texttt{CAMB} for $2 \leq \ell \leq 8000$.
    \item The suppressed lensing power spectrum $C_L^{\phi\phi,\mathrm{sup}} = A_{\rm lens}(L)C_L^{\phi\phi}$ from \texttt{CAMB} for $2 \leq L \leq 20\,000$.
    \item The partially lensed $TT$ power spectrum $\tilde{C}_\ell^{TT}$ from \texttt{CAMB} for $2 \leq \ell \leq 20\,000$.
    \item The quadratic estimator normalization $A_{TT}$ from \texttt{pytempura}\footnote{\url{https://github.com/simonsobs/tempura}}~\cite{Namikawa:2014yca} for $2 \leq L \leq 1208$.
    \item The expected lensing reconstruction bias $\NLkk$ from \eqnname~\eqref{eq:N1Kesden} with Monte Carlo integration in \texttt{cmbpix} for $2 \leq \ell \leq 1208$.
    \item The SCALE normalization $A_\Lcheck$ from \eqnname~\eqref{eq:SCALEApp/ALDef} with Monte Carlo integration in \texttt{cmbpix} for $2 \leq \Lcheck \leq 1989$.
    \item The SCALE observables $\left\langle \Psi_\Lcheck \right\rangle$ from \eqnname~\eqref{eq:SCALEApp/PsiExpected} with Monte Carlo integration in \texttt{cmbpix} for $2 \leq \Lcheck \leq 1989$.
\end{enumerate}
\begin{table}
    \centering
    \caption[Binning scheme for CMB observables]{Summary of binning for CMB observables.}
    \label{tab:Binning}
    \begin{tabular}{|c|c|c|c|}
        \hline
        \rowcolor[HTML]{E7F9FF} Observable & Multipole range & Bin width & $N_b$ \\
        \hline
        $\tilde{C}_\ell^{TT}$ & $2 \leq \ell \leq 31$ & 1 & 30 \\
        \rowcolor[HTML]{EFEFEF} $\tilde{C}_\ell^{TT}$ & $32 \leq \ell \leq 3002$ & 30 & 99 \\
        $C_L^{\kappa\kappa,\mathrm{rec}}$ & $2 \leq L \leq 1280$ & 71 & 18 \\
        \rowcolor[HTML]{EFEFEF} $\Psi_\Lcheck$ & $2 \leq \Lcheck \leq 1989$ & 71 & 28 \\
        \hline
    \end{tabular}
\end{table}
We compute the (partially) lensed CMB $TT$ power spectrum out to \texttt{lmax=20000} with \texttt{CAMB} for all $N_\mathrm{train}=8192$ sets of parameters in the training range, remembering to set \texttt{lens\_potential\_accuracy=8} for high-$\ell$ accuracy \cite{McCarthy:2022}. The high-$\ell$ regime is required for the calculation of the SCALE quantities shown \eqnname~\eqref{eq:SCALEApp/ALDef}-\eqref{eq:SCALEApp/PsiExpected}, which depend on the small-scale power spectra. 
We bin the $TT$ power spectra into band-powers following the Planck prescription \cite{Planck:2019nip}\footnote{\url{https://wiki.cosmos.esa.int/planckpla2015/index.php/CMB\_spectrum\_\%26\_Likelihood\_Code}}. 
The 30 largest-scale powers between $2 \leq \ell \leq 31$ are kept unbinned, and the powers between $32 \leq \ell \leq 3002$ are binned into 99 band-powers of width $\Delta\ell = 30$. The bin weights are given by: 
\\
\begin{equation} 
    w_{\ell_b \ell} = \frac{\ell(\ell+1)}{\sum_{\ell \in b} \ell(\ell+1)}, 
\end{equation}
with band-powers $\tilde{C}_{\ell_b}^{TT} = \sum_{\ell \in b} w_{\ell_b \ell} \tilde{C}_\ell^{TT}$ and bin centers $\ell_b = \sum_{\ell \in b} w_{\ell_b \ell} \ell$. The binning procedure is summarized in Table~\ref{tab:Binning}.

We also compute the CMB lensing power spectrum out to $L = 20\,000$, as the small-scale $C_L^{\phi\phi}$ is required for the SCALE calculations. 
We compute the expected $N_L^{(1),\phi\phi}$ for each set of cosmological parameters using a Monte Carlo integration method for \eqnname~\eqref{eq:N1Kesden} included in \texttt{cmbpix}. 
Since the filter weight $F_{TT}$ is dependent on our choice in QE filtering, we hold it fixed in our training spectra by setting the lensing response $C_\ell^{T\nabla T}$ and the total observed spectrum $C_\ell^{TT,\mathrm{obs}} = \tilde{C}_\ell^{TT} + N_\ell^{TT}$ at the fiducial cosmology parameters described in Table~\ref{tab:Fiducial}. The dependence of $N_L^{(1),\phi\phi}$ on the cosmology of the training space is through the suppressed $C_L^{\phi\phi,\mathrm{sup}}$, along with a contribution from the lensing response $C_\ell^{T\nabla T}$ in $f_{TT}$ (which, as stated below Eq.~\ref{eq:QEf}, reflects the expected effects of lensing on the temperature field). 
Finally, we also compute the quadratic estimator normalization $A_{TT}$ for every set of parameters in the training space with Eq.~\ref{eq:QENorm}. 
We wish to include information from conventional CMB lensing observables in our likelihood, so we save the expected `reconstructed' spectrum $C_L^{\kappa\kappa,\mathrm{rec}}$ as band-powers with a bin width of $\Delta L = 71$ between $2 \leq L \leq 1208$: 
\begin{equation}\label{eq:ReconstructedSpectrum}
    C_L^{\kappa\kappa,\mathrm{rec}} = \frac{A_{TT}^2}{A_{TT,\mathrm{fid}}^2} C_L^{\kappa\kappa,\mathrm{sup}} + \NLkk.
\end{equation}

The conversion between lensing potential and lensing convergence power is $C_L^{\kappa\kappa} = [L(L+1)]^2 C_L^{\phi\phi} / 4$, and $\mathcal{D}_L^{dd} = [L(L+1)]^2 C_L^{\phi\phi} / 2\pi$ is the default power spectrum returned by \texttt{CAMB}. 
The `reconstructed' spectrum reflects what the lensing convergence power spectrum $C_L^{\kappa\kappa}$ is expected to be for a given underlying cosmology given a chosen set of fiducial parameters for filtering (after subtracting $\Nokk$), hence the renormalization with respect to the fiducial normalization $A_{TT,\mathrm{fid}}$ in the first term. We further discuss the physical meaning of $C_L^{\kappa\kappa,\mathrm{rec}}$ in \S\ref{sec:SCALEApp/Likelihood}.

We compute expected SCALE observables for each set of parameters in our training space with Monte Carlo integration of \eqnname~\eqref{eq:SCALEApp/ALDef}-\eqref{eq:SCALEApp/PsiExpected} included in \texttt{cmbpix}. Similar to $\NLkk$, we hold the filter weights fixed at the fiducial cosmology: i.e., all factors of $C_\ell^{TT,\mathrm{fid}}$ and $C_\ell^{TT,\mathrm{obs}}$ in \eqnname~\eqref{eq:SCALEApp/ALDef}-\eqref{eq:SCALEApp/PsiExpected} are held fixed with respect to Table~\ref{tab:Fiducial}. The dependence of SCALE observables on the cosmological parameters in the training space (and by extension during parameter estimation) is through the lensing power spectrum $C_L^{\phi\phi}$ along with one factor each of $C_{\lL}^{TT}$ and $C_{|\vL-\vlL|}^{TT}$ in the numerator of the innermost integrals of \eqnname~\eqref{eq:SCALEApp/ALDef}-\eqref{eq:SCALEApp/PsiExpected}, which together reflect changes in the expected temperature trispectrum from lensing. 

Different parts of the high-$L$ lensing power are probed by altering the filtering scheme within SCALE: we include in the emulator the dependence of SCALE observables on the small-scale filter width $\Delta\lS$ and center $\overline{\lS}$, which is equivalent to altering the limits of the outer $\lS$ integral in \eqnname~\eqref{eq:SCALEApp/ALDef}-\eqref{eq:SCALEApp/PsiExpected}. As an example, $\lS \in [8\,000, 10\,000]$ corresponds to $\Delta\lS = 2\,000$ and $\overline{\lS} = 9\,000$. These `parameters' allow for flexibility when applying the likelihood for different SCALE data vectors constructed from the same map(s), but with different ranges of small-scale filtering. We further discuss this in \S\ref{sec:SCALEApp/Likelihood}. We save the expected $\Psi_\Lcheck$ for our whole training set as band-powers with a bin width of $\Delta \Lcheck = 71$ between $2 \leq \Lcheck \leq 1989$. The band-powers of all CMB observables that we include in our likelihood are summarized in Table~\ref{tab:Binning}.
\subsubsection{Emulator performance}\label{sec:SCALEApp/Emulator/Emulator/Performance}
The emulator we construct contains 4 hidden layers, each with 512 nodes, following the default configuration for a neural network emulator in \texttt{COSMOPOWER} \cite{COSMOPOWER:2022}. The input layer contains the 12 parameters in Table~\ref{tab:TrainingPriors}, and outputs are the combined 174 band-powers listed in Table~\ref{tab:Binning}. The relevant parameters for the emulator are converted to $\Omega_\mathrm{b}$, $\Omega_\mathrm{c}$, and $\ln(10^{10}A_s)$ as inputs rather than their counterparts in Table~\ref{tab:TrainingPriors}. 
We include 87.5\% of the full training set $N_\mathrm{train}=8192$ during training, with the remaining 1024 reserved for validation. The emulator takes approximately 1\,min~30\,s to train\footnote{Training time is quoted for a single NVIDIA GeForce RTX 3080 GPU with 10 GB of memory.}, and it only needs to be trained once for a given set of training spectra computed with a chosen cosmological model. The trained model can be saved and loaded for future use without retraining.

\begin{figure*}
    \centering
    \includegraphics[width=0.95\textwidth]{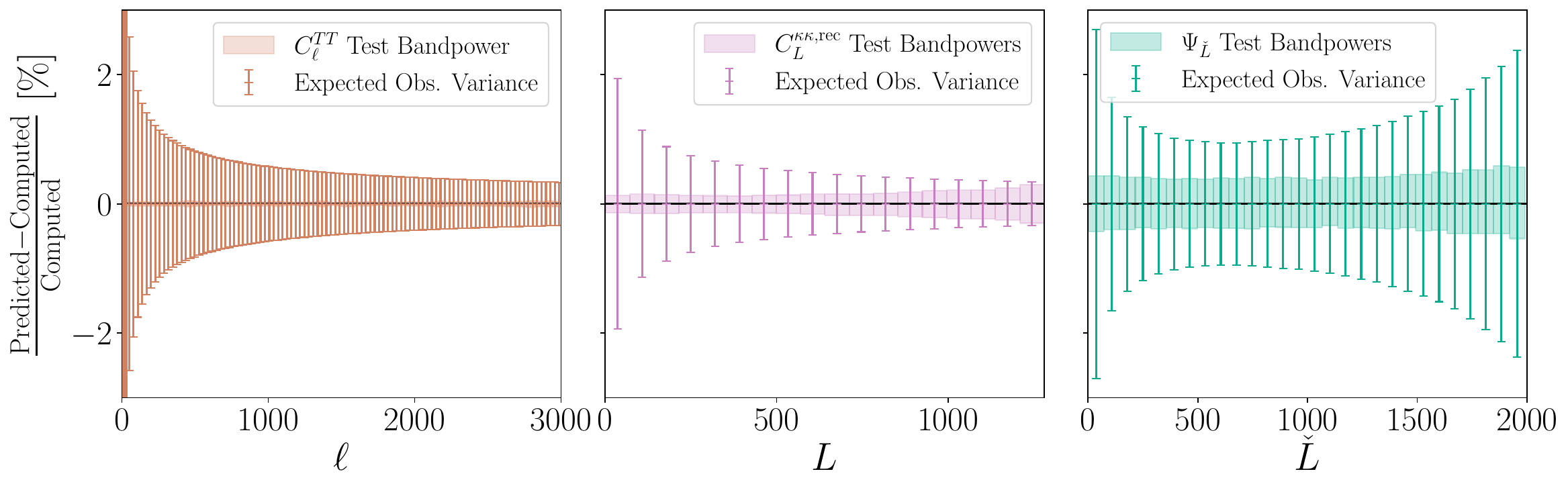}
    \caption[Validation of the emulator for CMB observables]{A validation of the emulator for our CMB observables. The \% error represented here is computed as the difference between the emulator prediction and the computed output (with \texttt{CAMB} for $\{ C_\ell^{TT}, C_L^{\kappa\kappa} \}$, and Monte Carlo integration with \texttt{CAMB} spectra for $\{ N_L^{(1),\kappa\kappa}, \Psi_{\check{L}} \}$) divided by the computed output. We perform training using 7168 out of a set of 8192 band-powers spanning a large range of cosmological parameters outlined in Table~\ref{tab:TrainingPriors}. We perform validation tests using the remaining 1024 sets of band-powers unseen by the emulator during training. The filled rectangles indicate the expected 68-percentile precision centered on the median, and the bin-to-bin scatter of emulator predictions is uncorrelated. The precision of each predicted $C_\ell^{TT}$ band-power is generally within $0.05\%$, and these rectangles are not visible on this $y$-scale. Error bars indicate the expected observational variance of band-powers, comparable to the diagonal of the covariance matrix underlying \figurename~\ref{fig:BaseCorrelation}. There is no significant bias from the emulator's predictions within our chosen range of cosmological parameter space, and the precision is generally better than the expected observational variance of all band-powers at our chosen level of noise.}
    \label{fig:EmulatorError}
\end{figure*}

We find that the emulator performs much more quickly than the original counterparts used to compute each set of band-powers, providing predictions of their respective observables in $\mathcal{O}(10^{-3}\,{\rm s})$. \figurename~\ref{fig:EmulatorError} shows that the lensed CMB $TT$ emulator is extremely precise, with a prediction error of $\lesssim 0.05\%$ per band-power. Similarly, \figurename~\ref{fig:EmulatorError} shows that the emulator predicts expected reconstructed band-powers with an error of $\lesssim 0.1\%$, and the SCALE observables with a precision of $\lesssim 0.5\%$ per band-power. The values of each band-power from the emulator are predicted independently and do not correlate with one another. This applies to their scatter as well; i.e., the bin-to-bin scatter of the emulator's predictions is uncorrelated. 
We attribute the lower precision of the reconstruction and SCALE band-powers to the inherent scatter of the input spectra computed with Monte Carlo integration related to the number of samples. We also find that the emulator is more accurate/precise when trained on the band-powers rather than their unbinned counterparts and then binning the full predicted spectra (a factor of $\sim 13.5\times$ more scatter for $TT$ band-powers and $\sim 1.5\times $ more scatter for the other band-powers). The scatter from predicting the full SCALE spectra multipole-by-multipole is at approximately the same level as the precision of the MC integration itself ($\sim 1\%$ before binning). Finally, the emulator exhibits an insignificant bias in its predictions, as the prediction errors effectively scatter around zero in \figurename~\ref{fig:EmulatorError}. A summary of the computational speedup provided by the emulator is provided in Table~\ref{tab:Speed}.
\begin{table}
    \centering
    \caption[Comparison of computation speed with and without emulators]{A comparison of computation speed for expected CMB observables. The computation time of $C_L^{\kappa\kappa}$ is comparable to that of $\tilde{C}_\ell^{TT}$. The computation times of $\NLkk$ and $A_\Lcheck$ are comparable to that of $\Psi_\Lcheck$. The emulator predicts all band-powers summarized in Table~\ref{tab:Binning} at once for a given set of parameters with a significant speedup for all observables.}
    \label{tab:Speed}
    \begin{tabular}{|l|c|}
        \hline
        \rowcolor[HTML]{E7F9FF} Computation & Time to evaluate [$\mathcal{O}({\rm s})$] \\
        \hline
        $\tilde{C}_{2<\ell<5{\rm k}}^{TT}$ (\texttt{CAMB}) & 1 \\
        \rowcolor[HTML]{EFEFEF} $\tilde{C}_{2<\ell<5{\rm k}}^{TT}$ (Emulator) & $10^{-3}$ \\
        $\Psi_{2<\Lcheck<2{\rm k}}$ (\texttt{cmbpix} mid-point) & $10^4$ \\
        \rowcolor[HTML]{EFEFEF} $\Psi_{2<\Lcheck<2{\rm k}}$ (\texttt{cmbpix} Monte Carlo) & 10 \\
        $\Psi_{2<\Lcheck<2{\rm k}}$ (Emulator) & $10^{-3}$ \\
        \hline
    \end{tabular}
\end{table}
\section{Simulations}\label{sec:SCALEApp/Sims}
We compute a suite of full-sky simulations of the lensed CMB temperature field with the \texttt{lenspyx}\footnote{\url{https://github.com/carronj/lenspyx}} package \cite{Lenspyx,Reinecke:2023}, which wraps around methods from \texttt{DUCC}\footnote{\url{https://gitlab.mpcdf.mpg.de/mtr/ducc}} (Distinctly Useful Code Collection) and allows for efficient and accurate lensing and de-lensing operations with spherical harmonics transforms \cite{Reinecke:2023}. A notebook with instructions to simulate the lensed CMB temperature and polarization is provided in the \texttt{lenspyx} repository. We choose to simulate maps with \texttt{HEALPix} resolution \texttt{NSIDE=8192}, as the \texttt{lenspyx} accuracy is good out to $\ell \approx 2\times\texttt{NSIDE}$. The simulations are constructed with the same fiducial cosmological parameters as shown in Table~\ref{tab:Fiducial}. We additionally apply the quadratic estimator with \texttt{so-lenspipe}\footnote{\url{https://github.com/simonsobs/so-lenspipe}} as well as SCALE with the following procedure:
\begin{figure*}
    \centering
    \includegraphics[width=0.95\hsize]{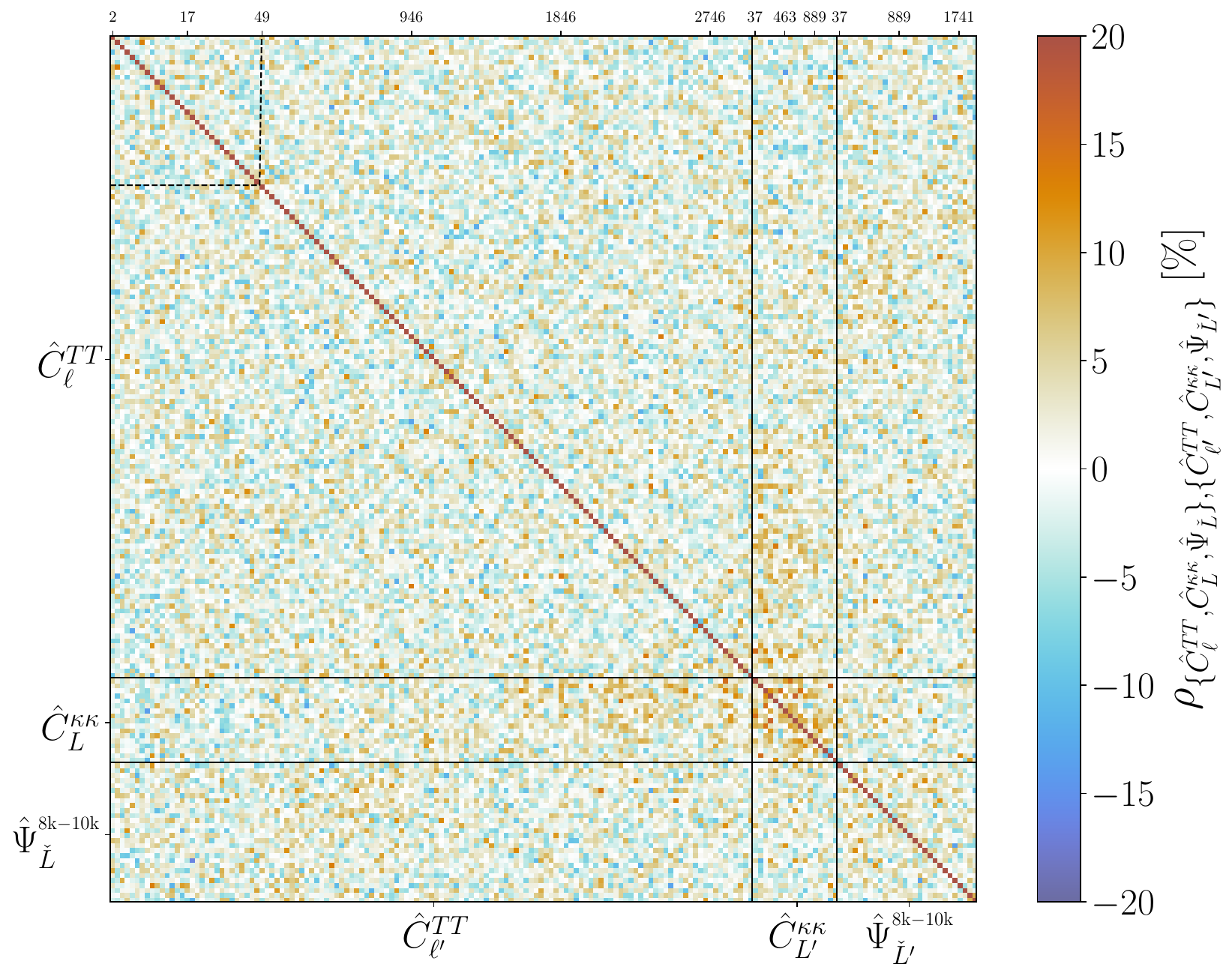}
    \caption[Correlation matrix for simulated CMB observable band-powers]{The correlation matrix between simulated CMB temperature, reconstructed lensing, and SCALE band-powers at the fiducial cosmology (see Table~\ref{tab:Fiducial}). The correlations are computed from 600 simulations. The band-powers are binned similarly to the Planck scheme for $\hat{C}_\ell^{TT}$ \cite{Planck:2019nip}: i.e., unbinned for $2 \leq \ell \leq 31$ (sectioned with dashed lines) and with width $\Delta\ell = 30$ for $32 \leq \ell \leq 3002$. The reconstructed lensing band-powers $\hat{C}_L^{\kappa\kappa,\mathrm{rec}}$ are binned with width $\Delta L=71$ for $2 \leq L \leq 1208$. SCALE band-powers $\hat{\Psi}_\Lcheck$ are binned with width $\Delta\Lcheck=71$ for $2 \leq \Lcheck \leq 1989$. Combinations of contributing components $\{ \hat{C}_\ell^{TT}, \hat{C}_L^{\kappa\kappa,\mathrm{rec}}, \hat{\Psi}_\Lcheck \}$ and their underlying covariance matrix are used in our likelihoods combining conventional CMB observables and SCALE band-powers.}
    \label{fig:BaseCorrelation}
\end{figure*}
\begin{enumerate}
    \item[A.] Generate a lensed temperature field, and estimate the total `observed' temperature power $\hat{C}_\ell^{TT}$.
    \begin{enumerate}
        \item[1.] Compute the fiducial unlensed CMB $TT$ power spectrum $C_\ell^{TT}$ and lensing potential power spectrum $C_L^{\phi\phi}$ with \texttt{CAMB} out to \texttt{lmax=20000} with \texttt{lens\_potential\_accuracy=8}.
        \item[2.] Generate spherical harmonic coefficients $T_{\ell m}$ and $\phi_{\ell m}$ for both an unlensed temperature $T$ using $C_\ell^{TT}$ and lensing potential $\phi$ field using $C_L^{\phi\phi}$ with \texttt{synalm}.
        \item[3.] Transform the lensing potential field into a spin-1 deflection field $\mathbf{d}$ with \texttt{lenspyx}'s \texttt{almxfl} method.
        \item[4.] Compute the lensed temperature field $\tilde{T}$ using the unlensed temperature $T$ and deflection $\mathbf{d}$ coefficients with \texttt{alm2lenmap}. This returns a lensed temperature field $\tilde{T}$ in map-space at \texttt{NSIDE=8192}.
        \item[5.] Generate spherical harmonic coefficients $N_{\ell m}$ for a Gaussian noise temperature field $N$ with \texttt{synalm} (noise parameters also in Table~\ref{tab:Fiducial}), convert to map-space with \texttt{alm2map}, and add to the lensed temperature field $T^\mathrm{obs} = \tilde{T} + N$.
        \item[6.] Convert the observed temperature field $T^\mathrm{obs}$ to spherical harmonic coefficients $T_{\ell m}^\mathrm{obs}$ using \texttt{map2alm}, and estimate the total `observed' $TT$ power spectrum $\hat{C}_\ell^{TT}$ with \texttt{alm2cl}, which is saved.
    \end{enumerate}
    \item[B.] Apply the quadratic estimator, and estimate the reconstructed lensing power spectrum $\hat{C}_L^{\kappa\kappa,\mathrm{rec}} = \hat{C}_L^{\kappa\kappa} - \RDNo$.
    \begin{enumerate}
        \item[1.] Apply an isotropic filter to $T_{\ell m}^\mathrm{obs}$ with the expected lensing response $C_\ell^{T\nabla T}$ such that $2 \leq \ell \leq 3000$ to get $T_{\ell m}^\mathrm{filt}$.
        \item[2.] Reconstruct the lensing potential field $\phi_{\ell m}^\mathrm{rec}$ from $T_{\ell m}^\mathrm{filt}$ with \texttt{so-lenspipe}, and estimate the reconstructed lensing power spectrum $\hat{C}_L^{\kappa\kappa}$ with \texttt{alm2cl}.
        \item[3.] Compute the realization dependent $\RDNo$ using the observed temperature power spectra $\hat{C}_\ell^{TT}$ with \texttt{so-lenspipe}.
        \item[4.] Save the total reconstructed lensing power spectrum $\hat{C}_L^{\kappa\kappa,\mathrm{rec}} = \hat{C}_L^{\kappa\kappa} - \RDNo$.
    \end{enumerate}
    \item[C.] Apply SCALE, and estimate the cross-spectrum between the large-scale temperature gradient, and the small-scale temperature gradient $\Psi_\Lcheck = A_\Lcheck C_\Lcheck^{\lambda\varsigma}$.
    \begin{enumerate}
        \item[1.] Apply a low-pass filter such that $0 < \lL < 3000$ along with a Wiener filter to the observed temperature field $T_\mathrm{obs}$ ($W_\lambda(\ell)$, shown below as \eqnname~\eqref{eq:SCALEApp/WienerFilter}) with \texttt{almxfl}. The product is a set of spherical harmonic coefficients, that when converted to map space with a spin-1 inverse transform \texttt{alm2map\_spin}, produces the two large-scale gradient components $[\nabla_{\theta} T_L, \nabla_\phi T_L / \sin\theta]$ that make up the $\lambda = (\nabla_{\theta} T_L)^2 + (\nabla_\phi T_L / \sin\theta)^2$ map of large-scale temperature gradient power required for one half of SCALE. Note that the filter is constructed with the theoretical spectra from Step A1 using \texttt{CAMB}, and follow the fiducial cosmology in Table~\ref{tab:Fiducial}.
        \begin{equation}\label{eq:SCALEApp/WienerFilter}
            W_\lambda(\ell) = 
            \begin{cases}
                \sqrt{\ell(\ell+1)} \frac{C_\ell^{TT}}{\tilde{C}_\ell^{TT} + N_\ell^{TT}} &,~ \ell < 3000 \\
                0 &,~ \ell \geq 3000
            \end{cases}  \, .
        \end{equation}
        \item[2.] Convert the $\lambda$ map into spherical harmonic space with \texttt{map2alm}.
        \item[3.] Apply a high-pass filter such that $\ell_{S,{\rm min}} < \lS < \ell_{S,{\rm max}}$ along with an inverse variance filter to the observed temperature field $T_\mathrm{obs}$ (shown below as \eqnname~\eqref{eq:SCALEApp/InvVarFilter}) with \texttt{almxfl}. The product is a set of spherical harmonic coefficients, that when converted to map space with a spin-1 inverse transform \texttt{alm2map\_spin}, produces the two small-scale gradient components $[\nabla_{\theta} T_S, \nabla_\phi T_S / \sin\theta]$ that make up the $\varsigma = (\nabla_{\theta} T_S)^2 + (\nabla_\phi T_S / \sin\theta)^2$ map of small-scale temperature gradient power required for the other half of SCALE. Note that the filter is constructed with the theoretical spectra from Step A1 using \texttt{CAMB}, and follow the fiducial cosmology in Table~\ref{tab:Fiducial}.
        \begin{equation}\label{eq:SCALEApp/InvVarFilter}
            W_\varsigma(\ell) = 
            \begin{cases}
                \sqrt{\ell(\ell+1)} \frac{1}{\tilde{C}_\ell^{TT} + N_\ell^{TT}} &,~ \ell_{1,{\rm min}} < \ell_1 < \ell_{1,{\rm max}} \\
                0 &,~ \mathrm{else}
            \end{cases}  \, .
        \end{equation}
        \item[4.] Convert the $\varsigma$ map into spherical harmonic space with \texttt{map2alm}.
        \item[5.] Estimate the cross spectrum $\hat{C}_\Lcheck^{\lambda\varsigma}$ between $\lambda$ and $\varsigma$ with \texttt{alm2cl}, and multiply by $A_\Lcheck$ to get $\hat{\Psi}_\Lcheck$ then save.
    \end{enumerate}
\end{enumerate}
\begin{figure}
    \centering
    \includegraphics[width=0.95\hsize]{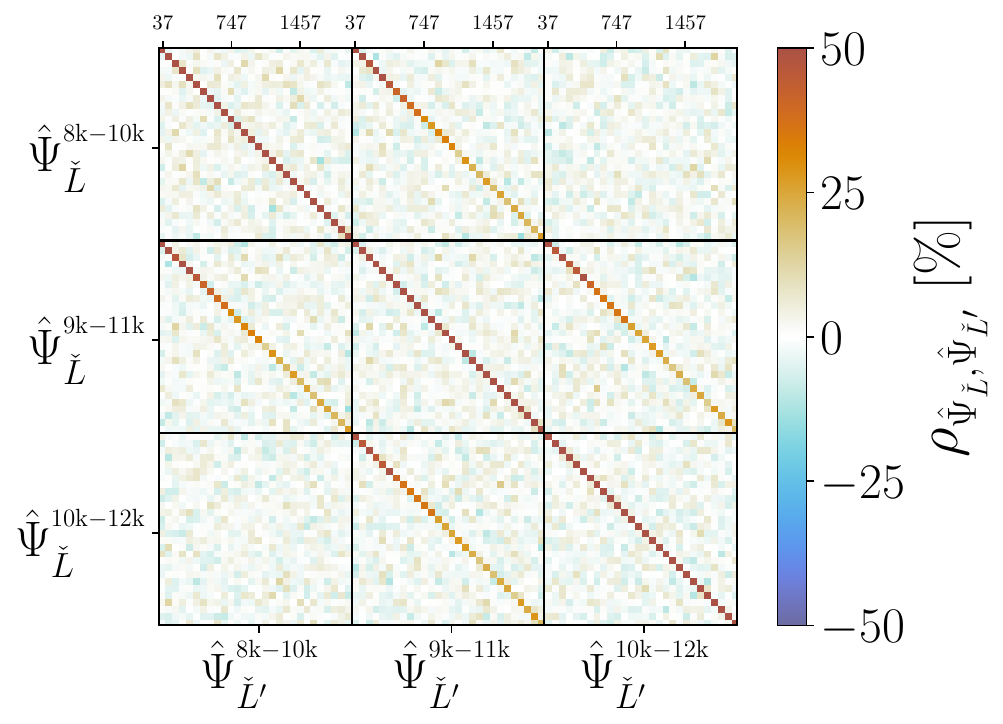}
    \caption[Correlation matrix between SCALE band-powers]{The correlation matrix between SCALE band-powers at the fiducial cosmology (see Table~\ref{tab:Fiducial}). The correlations between three applications of SCALE are depicted here, with shared large-scale filters $2 \leq \ell_L \leq 3000$ and different small-scale filters indicated by superscripts. The correlations are computed from 600 simulations. The strongest off-diagonal entries indicate that SCALE band-powers at the same $\Lcheck$ are approximately 20-50\% correlated between applications of small-scale filters which share half of their multipole coverage. Correlations between SCALE with the other $\lS$ filters $\{ \hat{\Psi}_\Lcheck^{9\k\mhyphen 11\k}, \hat{\Psi}_\Lcheck^{10\k\mhyphen 12\k} \}$ and the other CMB observables $\{ \hat{C}_\ell^{TT}, \hat{C}_L^{\kappa\kappa,\mathrm{rec}} \}$ are similar to those of $\hat{\Psi}_\Lcheck^{8\k\mhyphen 10\k}$ in \figurename~\ref{fig:BaseCorrelation}.
    }
    \label{fig:SCALECorrelation}
\end{figure}
We show in \S\ref{sec:SCALEApp/Likelihood} that, in principle, Part~C may be repeated on the same realization with various choices in the small-scale $\ell_{S,{\rm min}} < \lS < \ell_{S,{\rm max}}$ filter limits, and included in the same likelihood. We include three applications of SCALE for each realization: $8\,000 < \lS < 10\,000$, $9\,000 < \lS < 11\,000$, $10\,000 < \lS < 12\,000$, and apply their appropriate normalization $A_\Lcheck$ from \eqnname~\eqref{eq:SCALEApp/ALDef}. We repeat the above procedure for 600 simulations, and we save our CMB observables as band-powers of $\{ \hat{C}_\ell^{TT}, \allowbreak \hat{C}_L^{\kappa\kappa,\mathrm{rec}}, \hat{\Psi}_\Lcheck^{8\k\mhyphen 10\k}, \hat{\Psi}_\Lcheck^{9\k\mhyphen 11\k}, \hat{\Psi}_\Lcheck^{10\k\mhyphen 12\k} \}$ following Table~\ref{tab:Binning}. The correlation matrix between $\{ \hat{C}_\ell^{TT}, \allowbreak \hat{C}_L^{\kappa\kappa,\mathrm{rec}}, \hat{\Psi}_\Lcheck^{8\k\mhyphen 10\k} \}$ is shown in \figurename~\ref{fig:BaseCorrelation}. The correlation between SCALE observables and the other CMB observables is generally smaller than about $5\%$. This suggests that SCALE includes unique cosmological information from the small-scale CMB lensing. Note that this information is intentionally excluded from our quadratic estimator configuration because we showed in C24 that SCALE will outperform the QE at small-scales, and we wish to study how well SCALE can fulfill its role there.

We find that the SCALE observables exhibit very low levels of covariance between band-powers, as we found with the flat-sky simulations in C24. 
However, band-powers at the same $\Lcheck$ across small-scale filters with overlapping $\lS$ ranges (e.g., $\hat{\Psi}_\Lcheck^{8\k\mhyphen 10\k}$ and $\hat{\Psi}_\Lcheck^{9\k\mhyphen 11\k}$ for at the same $\Lcheck$) are highly correlated, with off-diagonal entries of approximately 20-50\%. 
This is shown in \figurename~\ref{fig:SCALECorrelation}, and it implies that applying SCALE with different small-scale filter ranges indeed characterizes different parts of the lensing power spectrum.

\begin{figure*}
    \centering
    \includegraphics[width=0.95\hsize]{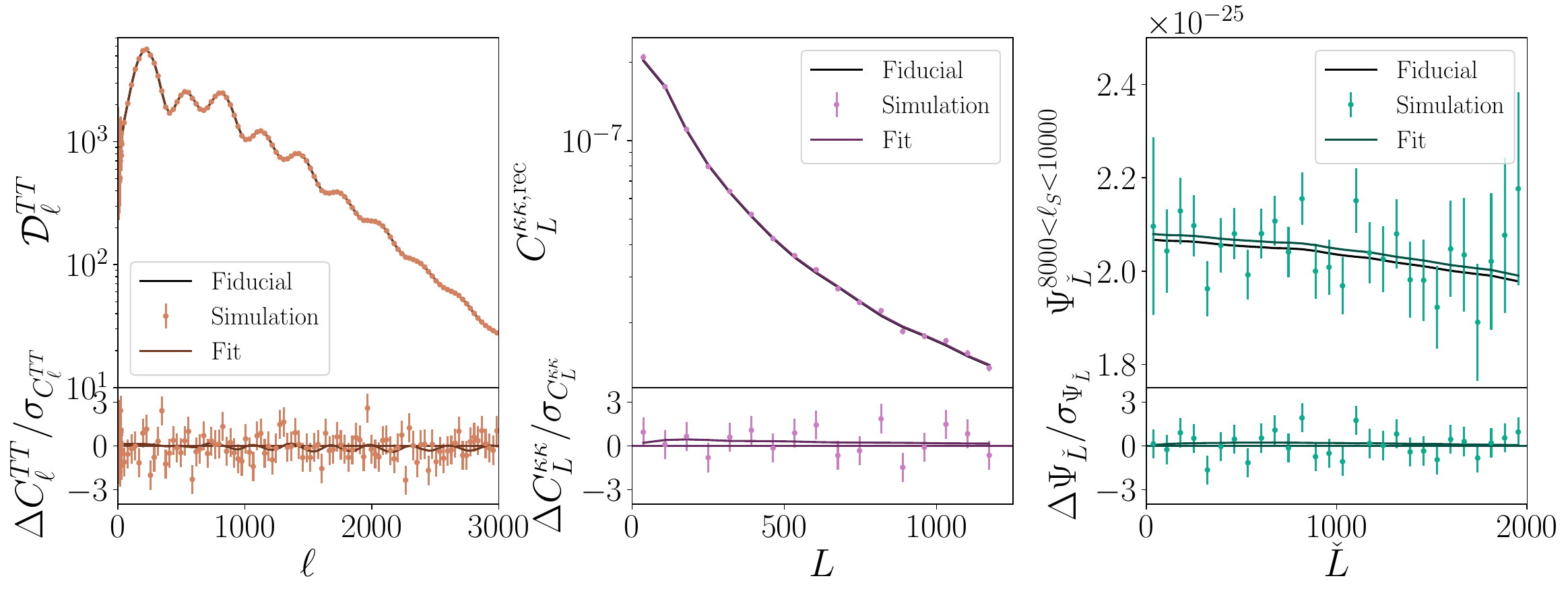}
    \caption[Comparison between theory and simulated observables]{A comparison between predicted CMB observable band-powers from the emulator at the fiducial cosmology with a set of band-powers computed from simulation. The simulated band-powers are overlaid with error-bars corresponding to the diagonal of \figurename~\ref{fig:BaseCorrelation}. Predicted band-powers at the best fit cosmology from combining all three observables in \S\ref{sec:SCALEApp/Results} are also shown. The residuals with respect to the fiducial band-powers are shown in the bottom panels. The `observed' band-powers from simulation match well with both the fiducial and best fit values.
    }
    \label{fig:DataFidTheory}
\end{figure*}

We additionally produce one more set of CMB observables with a separate realization following the above steps. This extra realization serves as the data vector that will be used in subsequent sections for cosmological inference, and is not included in the construction of the covariance/correlation matrices. A comparison of this realization's observable band-powers with the theoretical values at the fiducial cosmology is shown in \figurename~\ref{fig:DataFidTheory}. We produce one final set of CMB observables with a realization that includes a lensing suppression $C_L^{\kappa\kappa,\mathrm{sup}}$ in Step~A2 with $A_{\rm min} = 0.25$, and all other parameters following Table~\ref{tab:Fiducial}. This realization is used later to test SCALE's ability to constrain more exotic small-scale clustering phenomena.
\section{Constructing a Likelihood with SCALE}\label{sec:SCALEApp/Likelihood}
Consider a model defined by $\Lambda$CDM cosmology with the addition of one neutrino mass eigenstate parameterized with $m_\nu$. The general suppression of small-scale lensing due to the massive neutrino offers a simple test of SCALE's constraining power. Our initial vector of parameters is $\vec{\theta} = \{ m_\nu,\allowbreak\, \Omega_c,\allowbreak\, \Omega_b,\allowbreak\, \ln(10^{10}A_s),\allowbreak\, n_s,\allowbreak\, h,\allowbreak\, \tau \}$. The theoretical model that our emulator is trained on generally includes the lensing suppression at small scales, but for our initial analysis, we fix the relevant parameters $\vec{\theta}_\mathrm{sup} = \{ L_0, B, A_\mathrm{min} \}$ to their fiducial values in Table~\ref{tab:Fiducial}. In particular, setting $A_\mathrm{min} = 1$ effectively turns off the lensing suppression. There is an implicitly constrained parameter $\Omega_\Lambda$ such that $\Omega_c +\allowbreak \Omega_b +\allowbreak \Omega_\nu + \Omega_\Lambda = 1$. Our initial data vector $ \vec{d} = \{ \hat{C}_\ell^{TT}, \hat{C}_L^{\kappa\kappa,\mathrm{rec}},\allowbreak\, \hat{\Psi}_\Lcheck^{8\k\mhyphen 10\k} \}$ is composed of $TT$, QE reconstruction, and SCALE band-powers. We have constructed an emulator in \S\ref{sec:SCALEApp/Emulator} to predict the theoretical expected band-powers given a set of cosmological parameters: $\vec{t}(\vec{\theta})$. Finally, we have empirical estimates of the covariance between all band-powers of the data vector from our set of simulations in \S\ref{sec:SCALEApp/Sims}: $\mathbf{C}$. This allows us to construct a multivariate normal log-likelihood for the data vector $\vec{d}$ given a set of parameters $\vec{\theta}$:
\begin{align}
    \log p(\vec{d}|\vec{\theta}) &\sim \log \mathcal{N}(\vec{d}|\vec{t}(\vec{\theta}), \hat{\mathbf{C}}^{-1}) \nonumber\\
    &\sim - \frac{1}{2}(\vec{d}-\vec{t}(\vec{\theta}))^T\hat{\mathbf{C}}^{-1}(\vec{d}-\vec{t}(\vec{\theta})) \, .\label{eq:SCALEApp/Likelihood}
\end{align}
Note that the covariance $\mathbf{C}$ is estimated from $N_\mathrm{sims} = 600$ realizations, and the unbiased estimator for the inverse covariance must include the Hartlap factor \cite{Hartlap:2007}:
\begin{equation}\label{eq:SCALEApp/InvCov}
    \hat{\mathbf{C}}^{-1} = \frac{N_\mathrm{sims}-P-2}{N_\mathrm{sims}-1}\mathbf{C}^{-1} \, ,
\end{equation}
where $P$ is the number of band-powers included in the data vector. There are $P_{TT} = 129$ $TT$ band-powers, $P_\mathrm{QE} = 17$ QE band-powers, and $P_\mathrm{SCALE}=28$ band-powers for each application (with different small-scale $\lS$ filters) of SCALE. It is recommended that there is a minimum of realizations $N_\mathrm{sims} \gtrsim 2P$ to estimate the inverse covariance, which we have satisfied \cite{Hartlap:2007}.

The likelihood we apply compares the estimated QE reconstructed spectrum (when included) as 
\begin{align}\label{eq:QECompare}
    C_L^{\kappa\kappa,\mathrm{rec}} = \frac{A_{TT}^2}{A_{TT,\mathrm{fid}}^2}&C_L^{\kappa\kappa} + \NLkk \nonumber \\
    &\sim \hat{C}_L^{\kappa\kappa} - \RDNo = \hat{C}_L^{\kappa\kappa,\mathrm{rec}} \, ,
\end{align}
with $C_L^{\kappa\kappa,\mathrm{rec}}$ predicted by our emulator as a function of a given set of parameters $\vec{\theta}$, and $\hat{C}_L^{\kappa\kappa,\mathrm{rec}}$ estimated directly from each realization. 
In practice (e.g., with the ACT, SPT and Planck lensing analyses \cite{Planck:2015mym,Wu:2019hek, ACT:2023kun}), the QE-relevant quantities are directly compared to the expected CMB lensing power spectrum $C_L^{\kappa\kappa}$ as expressed in \eqnname~\eqref{eq:CLkkQE}. 
Our construction includes the dependence of $\NLkk$ on parameters directly on the theory side of the likelihood because the emulator allows us to quickly determine the expected reconstruction bias for each given set of parameters. This is in contrast to a more practical likelihood, which includes $\NLkk$ and its dependence on cosmological parameters as first-order corrections (with respect to the dependence of $\NLkk$ on $C_\ell^{TT}$ and $C_L^{\kappa\kappa}$) to the observed quantities due to challenges with repeatedly computing $\NLkk$ \cite{Planck:2015mym,Planck:2019nip,ACT:2023dou}. Our construction is equivalent to the practical likelihood in the limit of well-controlled, isotropic noise, and the usual treatment of $\NLkk$ is preferred for real data analysis (notably with the presence of foregrounds). 

We set broad and uniform priors $p_\mathrm{Uni}(\vec{\theta})$ for every parameter in $\vec{\theta}$ following Table~\ref{tab:TrainingPriors} except for $\tau$, for which we impose a Gaussian prior about the fiducial value (Table~\ref{tab:Fiducial}):
\begin{equation}\label{eq:SCALEApp/TauPrior}
    p(\tau) \sim \mathcal{N}(0.06,\sigma_\tau)\, ,
\end{equation}
where we choose either $\sigma_\tau = 0.007$ set by the value reported by \planck\ 2018 \cite{Planck:2018vyg}, or $\sigma_\tau = 0.002$ set by the cosmic variance limit \cite{LiteBIRD:2022cnt}. We also include an additional likelihood which includes forecasted constraints from Baryon Acoustic Oscillation (BAO) information from the 5-year survey of the Dark Energy Spectroscopic Instrument (DESI, \cite{DESI:2014}), which mainly constrains the matter density $\Omega_m$. We follow the steps in Appendix V of Ref.~\cite{Allison:2015} to construct a Fisher matrix $\mathbf{F}$ for BAO observables and the covariances between our other parameters, including $m_\nu$, $\Omega_c$, $\Omega_b$, and $h$.
The BAO log-likelihood is constructed as follows:
\begin{equation}\label{eq:SCALEApp/BAOLikelihood}
    \log p_{\mathrm{BAO}}(\vec{\theta}|\mathbf{F}) \sim -\frac{1}{2}(\vec{\theta} - \vec{\theta}_\mathrm{fid})^T \mathbf{F} (\vec{\theta} - \vec{\theta}_\mathrm{fid}) \, ,
\end{equation}
where $\vec{\theta}_\mathrm{fid}$ is the vector of fiducial parameters from Table~\ref{tab:Fiducial}. Our final posterior is then expressed by the following:
\begin{equation}\label{eq:SCALEApp/Posterior}
    p(\vec{\theta}|\vec{d}) \sim \mathcal{N}(\vec{d}|\vec{t}(\vec{\theta}), \hat{\mathbf{C}}^{-1}) \mathcal{N}(\tau|0.06,\sigma_\tau) p_\mathrm{BAO}(\vec{\theta}|\mathbf{F}) p_\mathrm{Uni}(\vec{\theta}) \, .
\end{equation}
\begin{figure*}
    \centering
    \includegraphics[width=0.96\textwidth]{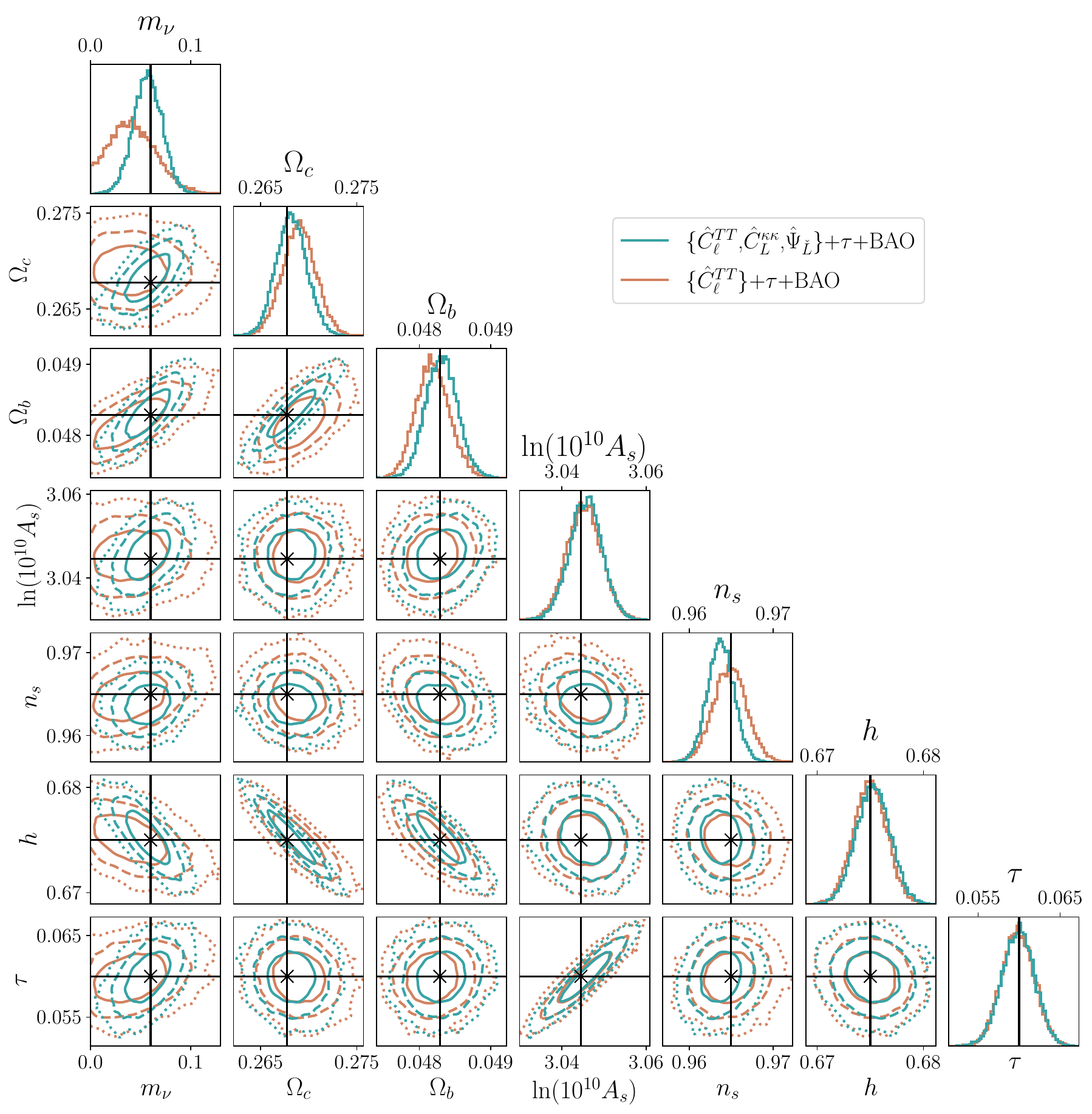}
    \caption[Full sampled posterior distributions for a simple cosmological model]{The sampled posterior distribution of the probabilistic model described in \eqnname~\eqref{eq:SCALEApp/Posterior} using the $TT$ band powers along with a cosmic variance $\tau$ prior $\sigma_\tau=0.002$, and BAO likelihood is shown in \textit{brown}. The resulting sampled posterior distribution with the addition of lensing information from both lensing reconstruction and SCALE is shown in \textit{green}. The contours indicate the 68\% (solid), 95\% (dashed) and 99.7\% (dotted) regions of a kernel density estimate for each 2-dimensional marginalized posterior. The fiducial values are indicated with black lines and $\times$ symbols. The addition of CMB lensing observables makes the difference in a detection of the minimum neutrino mass $m_\nu$ (see \figurename~\ref{fig:mnu}). We also see that the covariances between matter clustering related parameters such as $m_\nu$, $\Omega_c$ and $\Omega_b$ tighten up with the addition of lensing observables.}
    \label{fig:BasePosterior}
\end{figure*}
\begin{table*}
    \centering
    \caption[Summary of cosmological parameter constraints with and without SCALE]{Summary of cosmological parameter constraints from combinations of the observed \texttt{TT}, QE reconstructed, and SCALE band-powers, along with $\tau$ and BAO priors. Fiducial values from Table~\ref{tab:Fiducial} are also shown for comparison. The cosmic variance $\tau$ prior $\sigma_\tau^\mathrm{CV}=0.002$ \cite{LiteBIRD:2022cnt} is used here. Reported values are the median and 68\% credibility interval of each marginalized posterior. The median values of the posterior slightly shift around the fiducial values depending on the realization, but in general the shifts are comfortably within the 68\% credibility intervals. The widths of the 68\% regions do not change appreciably between realizations.}
    \label{tab:SummaryResults}
    \begin{tabular}{|l|c|c|c|c|c|}
        \hline
        \rowcolor[HTML]{E7F9FF} Parameter & Fiducial & $\hat{C}_\ell^{TT}$ & $\hat{C}_\ell^{TT}, \hat{C}_L^{\kappa\kappa,\mathrm{rec}}$ & $\hat{C}_\ell^{TT}, \hat{\Psi}_\Lcheck^{8\k\mhyphen 10\k}$ & $\hat{C}_\ell^{TT}, \hat{C}_L^{\kappa\kappa,\mathrm{rec}}, \hat{\Psi}_\Lcheck^{8\k\mhyphen 10\k}$ \\
        \hline
        $m_\nu$ [eV] & 0.06 & $0.041_{-0.023}^{+0.024}$ & $0.045_{-0.015}^{+0.015}$ & $0.066_{-0.015}^{+0.015}$ & $0.057_{-0.014}^{+0.014}$ \\
        \rowcolor[HTML]{EFEFEF} $\Omega_c$ & $0.2678$ & $0.2690_{-0.0017}^{+0.0017}$ & $0.2688_{-0.0016}^{+0.0016}$ & $0.2681_{-0.0015}^{+0.0016}$ & $0.2682_{-0.0016}^{+0.0016}$ \\
        $\Omega_b$ & $0.04829$ & $0.04818_{-0.0002}^{+0.0002}$ & $0.04820_{-0.0002}^{+0.0002}$ & $0.04829_{-0.0003}^{+0.0003}$ & $0.04831_{-0.0002}^{+0.0002}$ \\
        \rowcolor[HTML]{EFEFEF} $\ln(10^{10}A_s)$ & $3.045$ & $3.045_{-0.004}^{+0.004}$ & $3.045_{-0.004}^{+0.004}$ & $3.044_{-0.004}^{+0.004}$ & $3.046_{-0.004}^{+0.004}$ \\
        $n_s$ & $0.965$ & $0.965_{-0.002}^{+0.002}$ & $0.965_{-0.002}^{+0.002}$ & $0.966_{-0.002}^{+0.002}$ & $0.964_{-0.002}^{+0.002}$ \\
        \rowcolor[HTML]{EFEFEF} $h$ & $0.675$ & $0.675_{-0.002}^{+0.002}$ & $0.675_{-0.002}^{+0.002}$ & $0.675_{-0.002}^{+0.002}$ & $0.675_{-0.002}^{+0.002}$ \\
        $\tau$ & $0.06$ & $0.060_{-0.002}^{+0.002}$ & $0.060_{-0.002}^{+0.002}$ & $0.060_{-0.002}^{+0.002}$ & $0.060_{-0.002}^{+0.002}$ \\
        \hline
    \end{tabular}
\end{table*}
In principle, the data vector $\vec{d}$, covariance matrix $\mathbf{C}$, and theory vector constructed from emulators $\vec{t}(\vec{\theta})$ can contain any combination of $TT$, QE, and SCALE band-powers as long as all three are constructed consistently. We present in \S\ref{sec:SCALEApp/Results} results from several combinations of observable band-powers. We only consider one application of SCALE with $\ell_{S,{\rm min}} = 8\,000$ and $\ell_{S,{\rm max}} = 10\,000$ for our initial analysis. The model including a massive neutrino simply shifts the amplitude of the lensing potential power in the small-scale regime that we consider, so a single SCALE estimator is sufficient.

Including multiple applications of SCALE with different $\ell_{S,{\rm min}}$ and $\ell_{S,{\rm max}}$ in a single fit should allow for constraints on models which change the shape of the lensing power spectrum $C_L^{\kappa\kappa}$. To test this, we perform a separate analysis including the other two applications of SCALE in our likelihood with a full covariance combining \figurename~\ref{fig:BaseCorrelation} and \figurename~\ref{fig:SCALECorrelation}. A separate data vector from the realization in \S\ref{sec:SCALEApp/Sims} \textit{with} lensing suppression following Table~\ref{tab:Fiducial} with $A_\mathrm{min} = 0.25$ is used, and we allow the relevant parameters $\vec{\theta}_\mathrm{sup}$ to be free and sampled. We sample the logarithm of the decay rate parameter $\ln(B)$ to allow a larger dynamic range.

We construct and sample our probabilistic model with the Python implementation of Markov Chain Monte Carlo (MCMC) techniques in \texttt{emcee}\footnote{\url{https://github.com/dfm/emcee}} \cite{emcee:2013}. We choose to use \texttt{emcee} rather than more commonly-used software designed specifically for cosmology such as \texttt{CosmoMC}\footnote{\url{https://github.com/cmbant/CosmoMC}} \cite{CosmoMC:2002} or \texttt{cobaya}\footnote{\url{https://github.com/CobayaSampler/cobaya}} \cite{Cobaya:2019,Cobaya:2021} because it offers a simple way to construct log-probabilities with the added flexibility of allowing for the use of black-box functions in the model. The latter point is essential in order to use the emulator constructed in \S\ref{sec:SCALEApp/Emulator} at each step of the chain. We use 14 walkers, or chains if holding $\vec{\theta}_\mathrm{sup}$ fixed, and 20 walkers if $\vec{\theta}_\mathrm{sup}$ are free and sampled. Each chain is run for 50\,000 steps each. The first 1\,000 steps are discarded as burn-in steps that have yet to converge, and we further thin the chains by a factor of 50 to reduce the auto-correlation between samples. We find that the chains converge (satisfying the Gelman-Rubin ratio requirement $R < 1.1$) after approximately 1\,000 steps post-burn-in and before thinning. The results of each model are presented in \S\ref{sec:SCALEApp/Results}.
\section{Results}\label{sec:SCALEApp/Results}
\begin{figure*}
    \centering
    \includegraphics[width=0.95\hsize]{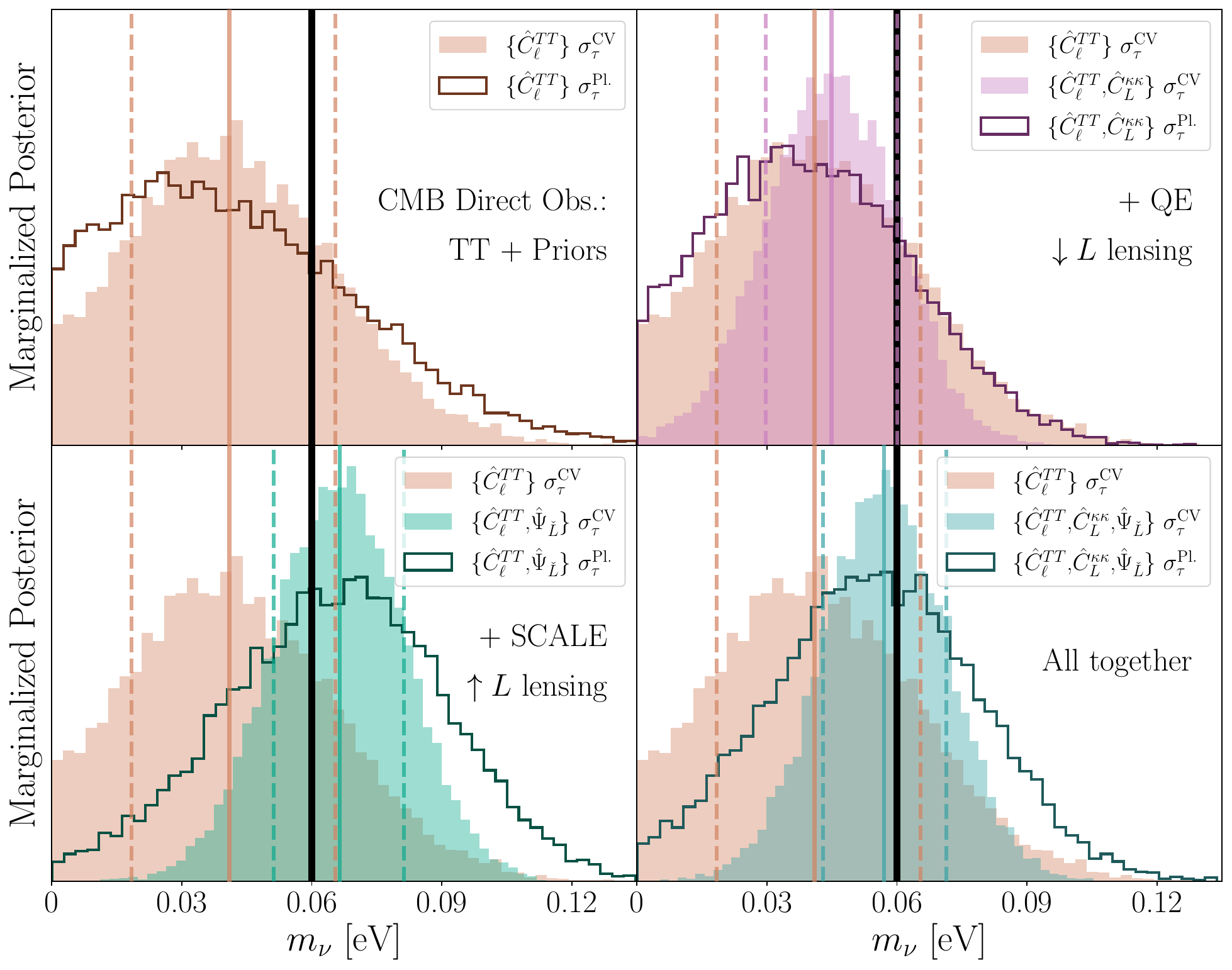}
    \caption[Marginalized posterior distributions for estimated neutrino mass]{The marginalized posteriors (with the cosmic variance $\tau$ prior $\sigma_\tau=0.002$) for $m_\nu$ with various combinations of CMB observables. The median of each result is indicated with solid vertical lines of their respective colours, and coloured, dashed lines indicate their 68\% credible intervals. A vertical black line indicates the fiducial value at $m_\nu = 0.06\,{\rm eV}$. The marginalized posteriors using the \planck\ 2018 $\tau$ prior $\sigma_\tau=0.007$ are shown in each respective panel with step histograms (medians and 68\% intervals not shown). With the condition of a precise $\tau$ estimate, the addition of lensing band-powers in general tightens the distribution of the marginalized posterior enough for a $4\sigma$ detection of minimum mass. The physical effect of neutrino mass is a mostly scale-independent change to the lensing amplitude, so this is true regardless of whether this lensing information comes from the QE reconstruction (top right), SCALE (bottom left), or both (bottom right).}
    \label{fig:mnu}
\end{figure*}

The results for our base model fits using the realization of observables presented in \figurename~\ref{fig:DataFidTheory} are summarized in Table~\ref{tab:SummaryResults} and \figurename~\ref{fig:BasePosterior}-\ref{fig:mnu}. The band-powers as predicted by our emulator for the best-fit including all three observables is also included in \figurename~\ref{fig:DataFidTheory}. We find that the size of the 68\% region for the best fit results do not change appreciably if we choose a different realization for the data vector. The center of the best fit can vary slightly between realizations, but the change is generally well within the 68\% range. The base model allows the $TT$ band-powers, combined with the BAO likelihood and $\tau$ prior, to constrain most of the parameters $\vec{\theta}$ to high precision in the presence of noise levels similar to that expected from CMB-S4~\cite{CMB-S4:2016ple,Abazajian:2019eic}. The exception is $m_\nu$, for which the $TT$-only model with either a \planck\ 2018 prior or a cosmic variance prior on $\tau$ is not able to detect (see \figurename~\ref{fig:mnu}). We also see in \figurename~\ref{fig:BasePosterior} and \figurename~\ref{fig:mnu} that the marginalized posterior for $m_\nu$ using only the $TT$ band-powers as observables causes the distribution to hit the edge of the prior at $0\,{\rm eV}$.

\begin{figure}
    \centering
    \includegraphics[width=0.95\hsize]{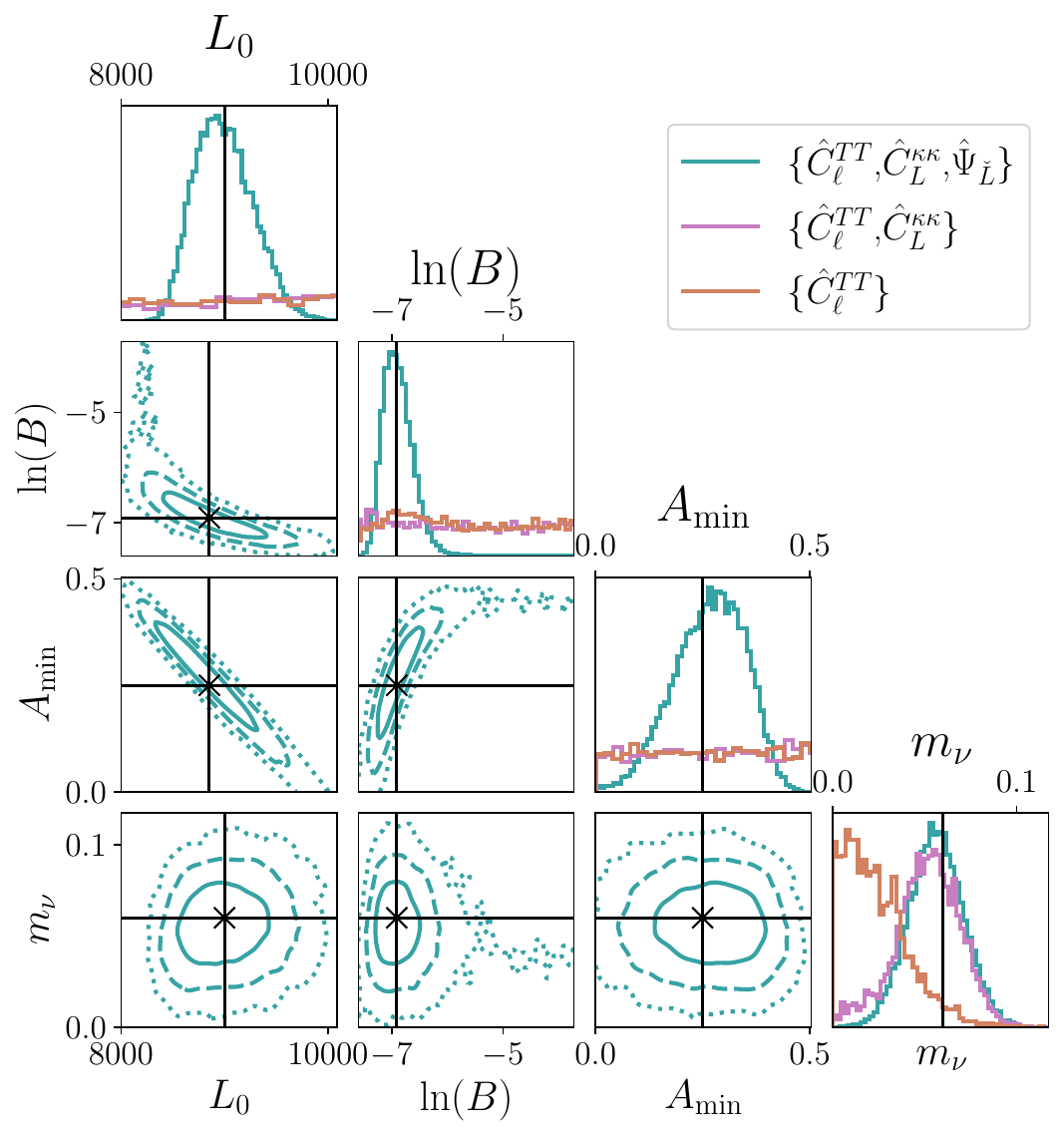}
    \caption[Sampled posterior distribution for a model with suppressed small-scale lensing]{The partially marginalized, sampled posterior for a simulation/model with suppressed lensing at small scales as depicted in \figurename~\ref{fig:LensSuppression}. All six $\Lambda$CDM parameters were also sampled with results similar to \figurename~\ref{fig:BasePosterior}, so they are not shown here. Including just a lensing reconstruction at large-scales provides sufficient information to constrain $m_\nu$. Only the full set of 3 SCALE band-powers (small-scales filtered for $ \lS \in \{ 8\k\mhyphen 10\k, 9\k\mhyphen 11\k, 10\k\mhyphen 12\k \}$) provides constraining information about this particular 3-parameter suppression model.}
    \label{fig:AlensCorner}
\end{figure}
\begin{table}
    \centering
    \caption[Summary of cosmological parameter constraints with a lensing suppression model]{Summary of cosmological parameter constraints on a model with small-scale lensing suppression from the observed \texttt{TT}, QE reconstructed, and SCALE band-powers, along with $\tau$ and BAO priors. Fiducial values from Table~\ref{tab:Fiducial} are also included. The cosmic variance $\tau$ prior $\sigma_\tau^\mathrm{CV}=0.002$ \cite{LiteBIRD:2022cnt} is used here. Reported values are the median and 68\% credibility interval of each marginalized posterior. The median values of the fit slightly shift around the fiducial values depending on the realization, but in general the shifts are comfortably within the 68\% credibility intervals. The widths of the 68\% regions do not change appreciably between realizations.}
    \label{tab:LensSupResults}
    \begin{tabular}{|l|c|c|}
        \hline
        \rowcolor[HTML]{E7F9FF} Parameter & Fiducial & $ \hat{C}_\ell^{TT}, \hat{C}_L^{\kappa\kappa,\mathrm{rec}}, \hat{\Psi}_\Lcheck^{8\k\mhyphen 10\k}, \hat{\Psi}_\Lcheck^{9\k\mhyphen 11\k}, \hat{\Psi}_\Lcheck^{10\k\mhyphen 12\k} $ \\
        \hline
        $L_0$ & 9000 & $8977_{-264}^{+309}$ \\
        \rowcolor[HTML]{EFEFEF} $10^3 B$ & 1.00 & $1.00_{-0.21}^{+0.34}$ \\
        $A_\mathrm{min}$ & 0.25 & $0.27_{-0.09}^{+0.08}$ \\
        \rowcolor[HTML]{EFEFEF} $m_\nu$ [eV] & 0.06 & $0.056_{-0.015}^{+0.015}$ \\
        $\Omega_c$ & $0.2678$ & $0.2678_{-0.0016}^{+0.0016}$ \\
        \rowcolor[HTML]{EFEFEF} $\Omega_b$ & $0.04829$ & $0.04835_{-0.0002}^{+0.0002}$ \\
        $\ln(10^{10}A_s)$ & $3.045$ & $3.042_{-0.004}^{+0.004}$ \\
        \rowcolor[HTML]{EFEFEF} $n_s$ & $0.965$ & $0.966_{-0.002}^{+0.002}$ \\
        $h$ & $0.675$ & $0.675_{-0.002}^{+0.002}$ \\
        \rowcolor[HTML]{EFEFEF} $\tau$ & $0.06$ & $0.060_{-0.002}^{+0.002}$ \\
        \hline
    \end{tabular}
\end{table}
The addition of SCALE into the data vector affects the parameters most sensitive to lensing (see \figurename~\ref{fig:BasePosterior}): $m_\nu$, $\Omega_c$, and $\Omega_b$. We see that the addition of SCALE alters the degeneracies between these parameters to become more constraining. Perhaps the most salient effect of including SCALE is the added ability to provide evidence for non-zero $m_\nu$ at $2.4\sigma$ with the \planck\ $\tau$ prior. The effect is more prominent if we swap the $\tau$ prior to the cosmic variance limit $\sigma_\tau = 0.002$ (see Table~\ref{tab:SummaryResults}), which is forecasted to be achievable with upcoming data from the LiteBIRD satellite mission \cite{LiteBIRD:2022cnt} or the CLASS ground-based survey \cite{Essinger-Hileman:2014pja}. In this case the detection jumps to a signal-to-noise of $4\sigma$ with the inclusion of SCALE. We have thus demonstrated that the addition of small-scale lensing information with SCALE provides extra constraining power for parameters which alter the lensing amplitude at $\ell \gg 3000$.

Finally, we present the results of our lensing suppression analysis in \figurename~\ref{fig:AlensCorner} and Table~\ref{tab:LensSupResults}. We find that the three sets of SCALE band-powers with different small-scale filtering regimes provides sufficient information to well-constrain the three parameters of our general lensing suppression model. This is the case even though the small-scale regimes we consider are overlapping, and technically share similar information about the small-scale lensing power spectrum. We find that the three lensing suppression parameters $\vec{\theta}_\mathrm{sup}$ are not well constrained if we perform the same analysis with a data vector consisting of only two of the applications of SCALE: for example with $\vec{d} = \{ \hat{C}_\ell^{TT}, \hat{C}_L^{\kappa\kappa,\mathrm{rec}}, \hat{\Psi}_\Lcheck^{8\k\mhyphen 10\k}, \hat{\Psi}_\Lcheck^{10\k\mhyphen 12\k} \}$. This is expected due to our three-parameter model requiring at least three measured amplitudes of the small-scale lensing power spectrum to constrain. In a similar vein, the same analysis with only conventional CMB observables $\{ \hat{C}_\ell^{TT}, \hat{C}_L^{\kappa\kappa,\mathrm{rec}} \}$ leaves the lensing suppression parameters $\vec{\theta}_\mathrm{sup}$ completely unconstrained, as they do not provide any information about the small-scale lensing power spectrum by construction.Therefore, with the choice of simulated data employed here, the SCALE observables provide unique constraining power in the small-scale lensing regime.
\section{Discussion and Conclusions}\label{sec:SCALEApp/Concl}
In this paper we explored applications of the SCALE estimator to cosmological parameter estimation, and how much information SCALE provides on top of more conventional methods for CMB lensing measurement. We originally developed SCALE in C24 as a novel estimator for the amplitude of CMB lensing power at small-scales $\ell \gg 3000$. The small-scale lensing regime has yet been untouched by conventional lensing reconstruction methods alone due to limits in instrument sensitivity and concerns with foreground contamination. We showed in C24 that SCALE will outperform conventional quadratic estimators in this regime for upcoming/future experiments in terms of the signal-to-noise of a lensing amplitude measurement. It is difficult to make direct comparisons between the efficacy of SCALE methodology and other small-scale lensing techniques  because SCALE produces neither a reconstruction of the lensing field nor a direct estimate of the lensing power spectrum. Comparisons can rather be made by substituting the small-scale lensing information (provided by SCALE in this paper) in a full likelihood constructed similarly to that presented and used in \S\ref{sec:SCALEApp/Likelihood}-\ref{sec:SCALEApp/Results} with a lensing power spectrum estimated with other small-scale lensing methods. We expect that small-scale foregrounds do not correlate directly with the large-scale CMB primary temperature field, and we reserve a study on the effects of foreground contamination on SCALE for future work. Using SCALE is quick and simple to implement because of its nature as the cross-spectrum of the same temperature map with two different filters applied. SCALE's outputs directly inform us about the CMB lensing power at small scales, but they do not estimate the underlying lensing field. In this work, we extended the study of SCALE by providing a framework for its application in a practical cosmological parameter estimation with other CMB observables. We further demonstrate that SCALE fills a beneficial niche by providing useful information that is complementary to well-understood and high-performing lensing reconstruction techniques applied to larger scales $L \lesssim 2000$. 

The effect of massive neutrinos on the lensing power spectrum is a nearly scale-independent decrease in amplitude, so it is a useful gauge for SCALE's effectiveness in comparison with more established QE reconstructions. We confirmed that the inclusion of SCALE observables provides sufficient information about the lensing amplitude at small-scales to provide significant evidence for non-zero $m_\nu$ with a CMB-S4-like experiment. In terms of $m_\nu$, the results with SCALE in the absence of a conventional lensing reconstruction with a QE (bottom left panel of \figurename~\ref{fig:mnu}) have a comparable performance to the results using a lensing reconstruction without SCALE (top-right panel of \figurename~\ref{fig:mnu}), with both measurements depending on a well-constrained $\tau$. This places SCALE in an interesting position for upcoming studies as a highly effective cross-check of standard CMB lensing estimation methods. This is particularly interesting given recent cosmological measurements of neutrino mass that prefer values smaller than expected from flavor oscillations, and even favor negative neutrino masses~\cite{Craig:2024tky,Wang:2024hen,Green:2024xbb,Naredo-Tuero:2024sgf,Jiang:2024viw}.

We further established SCALE's role in future cosmological analysis of small-scale lensing with a model including a phenomenological suppression of lensing at high $L \sim 10\,000$. The (mostly) scale-independent nature of the effects of massive neutrinos on CMB lensing means that including one application of SCALE with a single small-scale filter $\lS$ is sufficient for its estimation. We showed that including multiple applications of SCALE with different $\lS$ filters would allow for additional constraints on models which alter the shape of the lensing potential power spectrum $C_L^{\phi\phi}$, particularly at small-scales. The lensing suppression model we chose to focus on in this work was tailored to be easily identified with our choice/configuration of $\lS$ SCALE filtering. The generalization and optimization of SCALE filtering configurations is a natural path for future work. This will open a wide window of opportunity for SCALE to build on the foundation provided by conventional CMB analysis, and place constraints on exotic forms of dark matter or clustering models which are predicted to have non-trivial effects on the shape of the lensing potential power spectrum.

In summary, we:

\begin{itemize}
    \item Constructed a neural network emulator for the lensed $TT$, QE reconstruction, and SCALE band-powers. The emulator provides quick mapping from cosmological parameters to our expected observable band-powers at $\lesssim 0.5\%$ precision (\figurename~\ref{fig:EmulatorError} and Table~\ref{tab:Speed}).
    \item Presented a procedure to simulate a large sample of high-resolution (\texttt{NSIDE=8192}), full-sky simulations of the lensed CMB with the \texttt{lenspyx} package. We also presented a procedure to compute SCALE observables from these full-sky \texttt{HEALPix} representations, and SCALE observables from this procedure match well with the flat-sky results in C24.
    \item Developed a likelihood which includes SCALE in parameter estimation with conventional CMB observables accounting for covariance between the different observable spectra. We find that SCALE band-powers 
    $\hat{\Psi}_\Lcheck$ exhibit low levels of correlation with $\hat{C}_\ell^{TT}$ and $\hat{C}_L^{\phi\phi}$ band-powers, but they share strong correlations with band-powers (only at the same multipole $\Lcheck$) from other applications of SCALE with overlapping small-scale $\lS$ filters. 
    We applied this to a standard $\Lambda$CDM model with the addition of a massive neutrino $m_\nu$.
    \item Demonstrated that SCALE can directly provide constraining information in the estimation of parameters, such as $m_\nu$, which affect the amplitude of small-scale lensing beyond measurements of the lensed CMB power spectrum. 
    \item Constructed a phenomenological model (motivated by warm and/or fuzzy dark matter clustering models) for small-scale lensing suppression and demonstrated that multiple applications of SCALE with different small-scale $\lS$ filtering regimes provide sufficient information to constrain non-trivial modulation to the shape of the lensing power spectrum. This, together with the above result, establishes SCALE's role in future cosmological analyses by providing complementary information about small-scale lensing in addition to conventional lensing reconstruction methods at large-scales.
\end{itemize}

The upcoming era contains a lineup of highly sensitive CMB surveys that will map large fractions of the sky to unprecedented depths. SCALE and related methods will be critical tools to extract maximal information on the nature of dark matter and the evolution of structures in our Universe. 

\section*{Acknowledgments}
The authors would like to thank Kendrick Smith for early discussions that inspired this work. The authors would also like to thank Keir Rogers for help in getting started with cosmological emulators. 
Canadian co-authors acknowledge support from the Natural Sciences and Engineering Research Council of Canada (NSERC).  RH is supported by Natural Sciences and Engineering Research Council of Canada Discoavery Grant Program and the Connaught Fund.
JM is supported by the US~Department of Energy under Grant~\mbox{DE-SC0010129} and by NASA through Grant~\mbox{80NSSC24K0665}. AvE acknowledges support from NASA grants \mbox{80NSSC23K0747}, \mbox{80NSSC23K0464}, and ~\mbox{80NSSC24K0665}, and NSF grant \mbox{588167}.
Some computational resources for this research were provided by SMU’s Center for Research Computing.
The Dunlap Institute is funded through an endowment established by the David Dunlap family and the University of Toronto. 
The authors at the University of Toronto acknowledge that the land on which the University of Toronto is built is the traditional territory of the Haudenosaunee, and most recently, the territory of the Mississaugas of the New Credit First Nation. They are grateful to have the opportunity to work in the community, on this territory.
Computations were performed on the SciNet supercomputer at the SciNet HPC Consortium. SciNet is funded by: the Canada Foundation for Innovation; the Government of Ontario; Ontario Research Fund - Research Excellence; and the University of Toronto. 

\bibliographystyle{utphys}
\bibliography{SCALEbib} 

\end{document}